\documentclass[iop]{emulateapj}
\usepackage{natbib}
\usepackage{amsfonts}
\usepackage{amsmath}
\usepackage[usenames]{color}
\usepackage{ulem}
\usepackage{verbatim}
\usepackage{longtable}
\usepackage{array}
\usepackage{epsfig}
\newcommand\msun{M$_\odot$}
\newcommand\mzams{$M_{\rm ZAMS}$}

\shorttitle{Supernova Remnant Progenitor Masses in M31}
\begin{document}

\title{Supernova Remnant Progenitor Masses in M31}

\author{Zachary G. Jennings\altaffilmark{1},
Benjamin F. Williams\altaffilmark{1},
Jeremiah W. Murphy\altaffilmark{2},
Julianne J. Dalcanton\altaffilmark{1},
Karoline M. Gilbert\altaffilmark{1,4},
Andrew E. Dolphin\altaffilmark{3},
Morgan Fouesneau\altaffilmark{1},
Daniel R. Weisz\altaffilmark{1}}

\altaffiltext{1}{Box 351580, The University of Washington
  Seattle, WA 98195; zachjenn@uw.edu}
\altaffiltext{2}{Department of Astrophysical Sciences, Princeton University, Princeton, NJ 08544}
\altaffiltext{3}{Raytheon, 1151 E. Hermans Road, Tucson, AZ 85706; adolphin@raytheon.com}
\altaffiltext{4}{Hubble Fellow}

\begin{abstract}
Using Hubble Space Telescope (HST) photometry, we age-date 59
supernova remnants (SNRs) in the spiral galaxy M31 and use
these ages to estimate 
zero-age main sequence masses (\mzams) for their
progenitors.
To accomplish this, we create 
color-magnitude diagrams (CMDs) and employ CMD fitting to
measure the recent star formation history (SFH) of the regions
surrounding cataloged SNR sites.
We identify any young coeval population that likely produced the progenitor star,
then assign an age and uncertainty to that population. Application
of stellar evolution models allows us to infer the \mzams from this age. Because
our technique is not contingent on identification or precise location of the
progenitor star, it can be applied to the location of any known SNR.
We identify significant young star formation
around 53 of the 59
SNRs and assign progenitor masses to these, representing 
a factor of $\sim$2 increase over currently measured progenitor masses. 
We consider the remaining 6 SNRs as either probable Type Ia candidates or the result
of core-collapse progenitors that have escaped their birth sites.
In general, the distribution of recovered progenitor masses is bottom
heavy, showing a paucity of the most massive stars. If we assume a
single power law distribution, $dN/dM \propto M^{\alpha}$, we find
a distribution that is steeper than a Salpeter IMF
($\alpha=-2.35$).  In particular, we find values of $\alpha$
outside the range $-2.7 \ge \alpha \ge -4.4$ to be inconsistent
with our measured distribution at 95\% confidence. If instead
we assume a distribution that follows a Salpeter IMF up to some
maximum mass, we find that values of $M_{Max} > 26$ are inconsistent
with the measured distribution at 95\% confidence. In either scenario, the data
suggest that some fraction of massive stars may not explode. The result is preliminary
and requires more SNRs and further analysis. In addition, we use our
distribution to estimate a minimum mass for core collapse
between 7.0 and 7.8~\msun.
\end{abstract}


\section{INTRODUCTION}
Both theoretically and
observationally, core-collapse supernova
(CCSN) explosions are linked with the death of massive stars.
However, because precise measurements of CCSN progenitor masses are
scarce, the mapping between explosion scenario and the
progenitor mass  distribution is less clear.
The most common method of
progenitor mass determination is direct imaging, in which one
identifies the progenitor star in multi-band pre-explosion imaging,
fits a spectral energy distribution to the photometry, and
verifies the star is gone once the SN has faded.
This methodology is ideal given appropriate data, but it is limited to
contemporary SNe that have pre-explosion Hubble Space Telescope (HST) or deep
ground-based imaging.
As a result, only $\sim$25 SNe have any constraint 
on their progenitor masses, and half of these
are only upper limits \citep{smartt2002,vandyk2003a,
vandyk2003b,smartt2004,maund2005,hendry2006,li2005,li2006,li2007,
smartt2009,smartt2009b,galyam2007,galyam2009,smith2011b,maund2011,vandyk2011,
fraser2012,vandyk2012b,vandyk2012a}. A significant
increase in the number of progenitor masses could allow us a powerful
window into further understanding supernovae.

Direct imaging has created a prototypical picture
of massive star death. Type II-P SNe are assumed to be
created by red supergiant (RSG) stars with
intact hydrogen envelopes, while more exotic SNe may be created by
higher mass stars. However, a number of questions remain to be answered
regarding supernova physics. \citet{smartt2009} identify what they term to
be the `red supergiant problem,' an observed lack of progenitors of \mzams
$\sim$16-30~\msun that we would expect to explode as Type II-P SN. A variety
of channels have been proposed to account for these missing explosions,
including direct black hole formation or the stars exploding as different
SN types. Binarity of the progenitor system can also play a significant role
in SN explosions. \citet{smith2011} show that numbers of Type Ibc and
IIb explosions are significantly underestimated if single-star explosions
are assumed to be the only SN channel.

These issues are difficult to address
without further progenitors to analyze. Unfortunately, while direct-imaging
of the progenitor star would be ideal, there are constraints
that limit the frequency with which it may be applied.
First, direct imaging is limited by the
SN rate: roughly one per century in a large spiral 
galaxy \citep{cappellaro1999}. Second, direct imaging requires that the host
galaxy be close enough to resolve the progenitor star,
an effective limit of $\sim$20 Mpc \citep{smartt2009b}. 
Third, preexisting sub-arcsecond images of the site must exist, 
necessitating archival HST images with either the Wide Field
Planetary Camera 2 (WFPC-2) or the Advanced Camera for Surveys (ACS).
Finally, even with the existence of appropriate images, there is no guarantee that the precursor will be 
identified. For example, \citet{smartt2009} summarized 20 progenitor detections
in Type II-P explosions. Only five progenitors were directly observed,
and an additional two which fell on compact star clusters were
well constrained. The remainder had no detection, but were given upper 
bounds based on the maximum luminosity with which they could escape 
detection (see also \citet{smartt2003,vandyk2003,maund2005a}).

In short, while direct imaging is ideal given appropriate data, it suffers from prerequisites
that limit the number of opportunities in which it may be applied. It is clear that
development of an independent and complementary technique would be of interest.  

In this paper we make use of an alternative technique, stellar population analysis,
to infer progenitor masses. The technique involves examining a color-magnitude diagram (CMD)
of the surrounding
population of stars to infer an age and mass for the progenitor star. The advantage
of our methodology is that we are not reliant on individual identification and photometry
of the specific progenitor star, leaving us free
to apply our method to cataloged supernova remnants (SNRs)
in addition to directly observed
SNe. This ability drastically increases the number
of progenitor masses we may find,
allowing us to make a more complete measurement of
the underlying distribution of progenitor masses.
Our method offers no direct way to probe the binarity of the system,
which is relevant given the role binarity likely plays in certain
CCSN scenarios.  However, our estimate of the age of the surrounding stellar population
is not affected by the particular details of a given progenitor system.
Our method also offers no way to determine the type of CCSN explosion
we are observing, so for SNR analysis we are not able to provide a direct
link between SN explosion type and progenitor mass. However, we will
be able to comment on the ranges of ZAMS
masses that are producing CCSN explosions. 

Stellar population analysis has been used many times
historically to analyze the characteristics of progenitors.
Many groups have carried out age dating of 
stellar clusters coincident with supernovae and derived
corresponding stellar masses \citep{efremov1991,
walborn1993,panagia2000,barth1996,vandyk1999,maiz2004,wang2005,
vinko2009,crockett2008}. \citet{Badenes2009} examined the 
SFH map of the LMC published by \citet{harris2009}. They 
estimate ages and masses for the four most recent SNR 
produced by CCSNe in the LMC. In \citet{gogarten2009}, we 
employed a technique identical to the one used in this paper to
estimate the age of the progenitor of NGC 300 OT2008-1. We 
found that the star was likely to have formed 8-13 
Myr ago, corresponding to a \mzams~of 12-17~\msun. 
Finally, in \citet{murphy2011} we applied the same 
technique as this paper to archival images of SN 2011dh, 
finding a most likely age of 17$^{+3}_{-4}$~Myr and 
a \mzams~of 13$^{+2}_{-1}$~\msun. Clearly
there is a long list of successful applications 
of stellar population analysis leading to
constraints on progenitor physical properties.

In this paper we examine 59 SNRs in the spiral galaxy 
M31, listed in Table~\ref{tab_snr}. In \S2 we outline the methods used 
to determine star formation histories (SFHs), ages, and masses for the progenitor 
stars. In \S3 we present the results of the analysis. We find that
53 SNR display recent star formation, and we derive masses for these.
We also present four example cases that are representative
of the overall sample. Finally, in \S4 we examine the distribution
of progenitor masses. We search for indications of a minimum mass, which we
find to be between 7.0 and 7.8~\msun, and note that the distribution is
more bottom-heavy than a Salpeter IMF distribution.

\section{METHODOLOGY}
The basic procedure for deriving a mass estimate for each progenitor
is as follows. We first
use coordinates from three SNR catalogs
as explained in \S2.2, which we assume are accurate
to within a few arcseconds and
unbiased to any one type of SN. We then search for
any appropriate HST fields that contain the SNR in
question. We perform photometry on the stars in each
field, and create CMDs of the region within $\sim$50 pc
of the SNR coordinates, assuming a distance modulus of
$m-M=24.47$~\citep{stanek1998}. We use CMD fitting to measure
the SFH of each region. For each region that displays
recent SF, we use the age of this recent SF to associate
an age with the progenitor star. Finally, we apply stellar
evolution models to convert the age of the progenitor
to the \mzams\ of the progenitor.

\subsection{Key Assumptions}
Our method requires several assumptions to produce results. 
Above all, we must assume that some of the stars surrounding the SNR are in
fact coeval with the progenitor star. There are several pieces of 
evidence suggesting this assumption is reasonable. First,
90\% of stars form in clusters with sizes of 
$\sim$1 pc \citep{lada2003}. These stars are expected to stay spatially 
associated for timescales greater than the lifetimes of CCSNe progenitors,
even for dissolving, unbound clusters. Second, 
theoretical predictions are that stars will stay spatially 
associated on scales of $\sim$100 pc for 100 Myr \citep{bastian2006}. 
Observational constraints support this \citep{gogarten2009b}, as do simulations
by \citet{eldridge2011} who found that $\sim$85\% 
of CCSNe will explode within 100 pc of their birth site.
In addition, even if a progenitor would travel farther, we would still 
expect to see a fraction of the coeval young stellar population which is
sufficient for age dating. 
The precise number of interest is the fraction of young stars that will stay
within a certain distance in the lifetime of a core collapse supernova progenitor.
The question is a complex one, and we know of no research that quantifies
it precisely at this time. We conclude that it is a
reasonable assumption that most of the young stellar population
around our observed SNRs is coeval with their progenitors.  
Based on the analysis in \citet{gogarten2009}, we adopt a value
of $\sim$50 pc for the radius of our star selection annulus.

Our method does not assume any information about the type of SN that exploded.
We assume that SNRs associated with recent star formation are CCSNe, but do
not distinguish individual subtypes. For CCSNe, the particular type
of supernova has no effect on our mass determination process.
However, the progenitors of thermonuclear Type Ia SNe are likely
to arise from older stellar populations which may vary considerably
more in age than the coeval populations of massive stars.
While there exists considerably discussion in the literature as to
the precise nature of Type Ia progenitor systems and their
distribution in age (see \citet{maoz2011} and references therein),
we would expect to see
Type Ia progenitors ranging from ages of a few hundred Myr to
several Gyr. Since these timescales are comparable to or significantly
greater than the dynamical timescale of M31, our methodology of
identifying and analyzing coeval stars is
ineffective at measuring the precise ages of Type Ia progenitors.

Our method does have the risk of misinterpreting a Type
Ia SNR as a CCSNR. In their volume limited sample (D $<$ 60 Mpc),
\citet{li2011} find that Type Ia SN compose
24\% of observed SN. However, there are several qualitative
reasons as to why one would expect a smaller Ia/CCSN fraction
in our survey. 

First, there is some discussion as to whether
such a volume-limited survey may underestimate the fraction
of CCSN due to their faintness compared to Type Ia SN (see discussions
in \citet{thompson2009,horiuchi2009,horiuchi2011}). This 
potential bias suggests
that 24\% could be considered an upper limit. In addition, the areas of
M31 with greater HST coverage are primarily star forming regions, where we
would naively expect CCSN to be more
common, increasing the relative fraction of CCSN.
Thus the actual fraction of Type Ia SNR we
expect in our sample is probably somewhat less than
24\%. 

Moreover, it is likely that some Type Ia SNR can be identified
by their lack of an associated young stellar population.
\citet{Badenes2009} examined four Ia sites in the LMC and 
found that three displayed a significant lack of recent star formation.
Out of the 59
examined, we found 6 SNR displayed no recent SF in their surrounding stellar populations.

Finally, we have examined what effect additional hidden Ia contamination
will have on our data. The procedure we used is explained in
\S4.3. We found that additional Ia contamination
up to a total fraction of $\sim$25\%
has very little effect on the overall distribution observed in the sample.

To summarize, we argue that there are several qualitative reasons
to assume that the effective Ia fraction in our sample is smaller than
24\%. In addition, our method offers a means to identify and remove
some SNR that are likely Ia progenitors. Finally, our overall results
are largely insensitive to inclusion of additional Ia contamination
up to the 24\% observed in the volume-limited survey from \citet{li2011}.

Another possibility for SNRs with no coincident young SF is that
they are the result of high-velocity progenitor stars that
have left their birth sites. As a test of this possibility,
we examined any site with no observed recent SF in GALEX FUV
data. We found that all 6 sites with no recent SF are still
relatively close ($\sim$50 to 100 pc) to GALEX FUV sources. If we assume that
the progenitor star resulted from the most nearby FUV source for each zero-SF SNR,
then a 50 Myr old star would only require velocities of a few
km/s to reach the SNR site from the FUV source site, well within
reasonable velocities for runaway stars. Higher mass
stars would of course require higher velocities to leave the
FUV source in their shorter lifetimes. This analysis suggests that
our study is unable to differentiate between Type Ia SNR and SNR
resulting from high-velocity stars. We note, however, that a null-result
for progenitor mass will not affect the distribution of progenitor masses
we measure. It will simply reduce our effective sample size. 

Finally, our method is highly contingent on the accuracy 
of the stellar evolution models used to model our observed CMDs.
Because the models are theoretical, they generally do not have 
easily quantifiable uncertainties. The wide array of available model sets
tend to systematically differ in bolometric luminosity and temperature,
but are consistent within about 0.2 mag in bolometric luminosity and 0.02 dex in log temperature.
We therefore quantify the uncertainty of our results due to the models used by including random
shifts in the bolometric luminosity and temperature of the models
as part of our Monte Carlo (MC) tests (see also \citet{dolphin2002}). We detail the process further
in \S2.5.

\subsection{Data and Photometry}
There are only a few extensive catalogs of SNRs in M31. We
use the catalogs of \citet{braun1993}, \citet{magnier1995},
and \citet{williams1995}. All three make identifications based on [SII]-to-H$\alpha$
ratios, and \citet{magnier1995} make additional use of morphology and 
OB star associations. \citet{magnier1995} identify three
confidence levels based on approximate levels of contamination,
although they note that the confidence levels are somewhat
subjective. We only perform analysis on their
first and second confidence candidates. The catalogs are not
mutually exclusive, and the coverage areas of the three catalogs
overlap in several areas, causing several remnants to be identified
in both catalogs. To avoid treating a double-identified
remnant as two separate remnants, we identify all remnants from
two separate catalogs
in which the 50 pc areas overlap and only 
select coordinates from one catalog. In these duplicate cases, we adopted the coordinates
first from \citet{braun1993}, then from \citet{magnier1995} if \citet{braun1993} did not identify
a remnant in that position. On occasion, a single catalog identified two remnants
in which these 50 pc regions overlapped, but we still performed analysis on both candidates
in such a case. In addition, we cross-referenced
the SNR catalogs with \citet{sasaki2012}
and eliminated two, K567 and K884, which were found to not be SNR based on their
optical/X-ray properties.

For our analysis, 
we require that the stellar population surrounding the SNR 
be imaged by either the Advanced Camera for Surveys (ACS) 
or the Wide-Field Planetary Camera 2 (WFPC2) instruments 
on HST in at least two broadband filters.
We also require our 50\% completeness
limits (see \S2.3) for a given CMD to be at least 24.5 magnitude
in F475W, F555W, or F606W,
whichever represents the blue filter (all qualifying fields
of sufficient depth were either F475W, F555W, or F606W vs. F814W).
The location of the MS turnoff is very similar in color in these filter sets,
so adopting a similar magnitude cut in all three is reasonable.

Our comparisons between SNRs imaged multiple times at varying depths
indicate that simply reaching the MS turnoff for a given CMD
is insufficient to correctly determine the age. CMD fitting
is highly sensitive to densities of stars at given ages, requiring a
well-sampled stellar population at the age of interest.
In general,
we found that our methodology applied to shallower data
had a tendency to miss older bursts of SF which still may
have resulted in CCSN.
The available 2-filter
broadband data for M31 happens to be distributed in two groups, including many shallow
fields (blue depth $<23$) and many deep fields (blue depth 
$>25$), with very few in between. The choice of our depth cut
causes us to reject five SNR that have only been imaged with
shallow WFPC-2 data.

\begin{figure}
  \centering
   \includegraphics[width=0.8\textwidth]{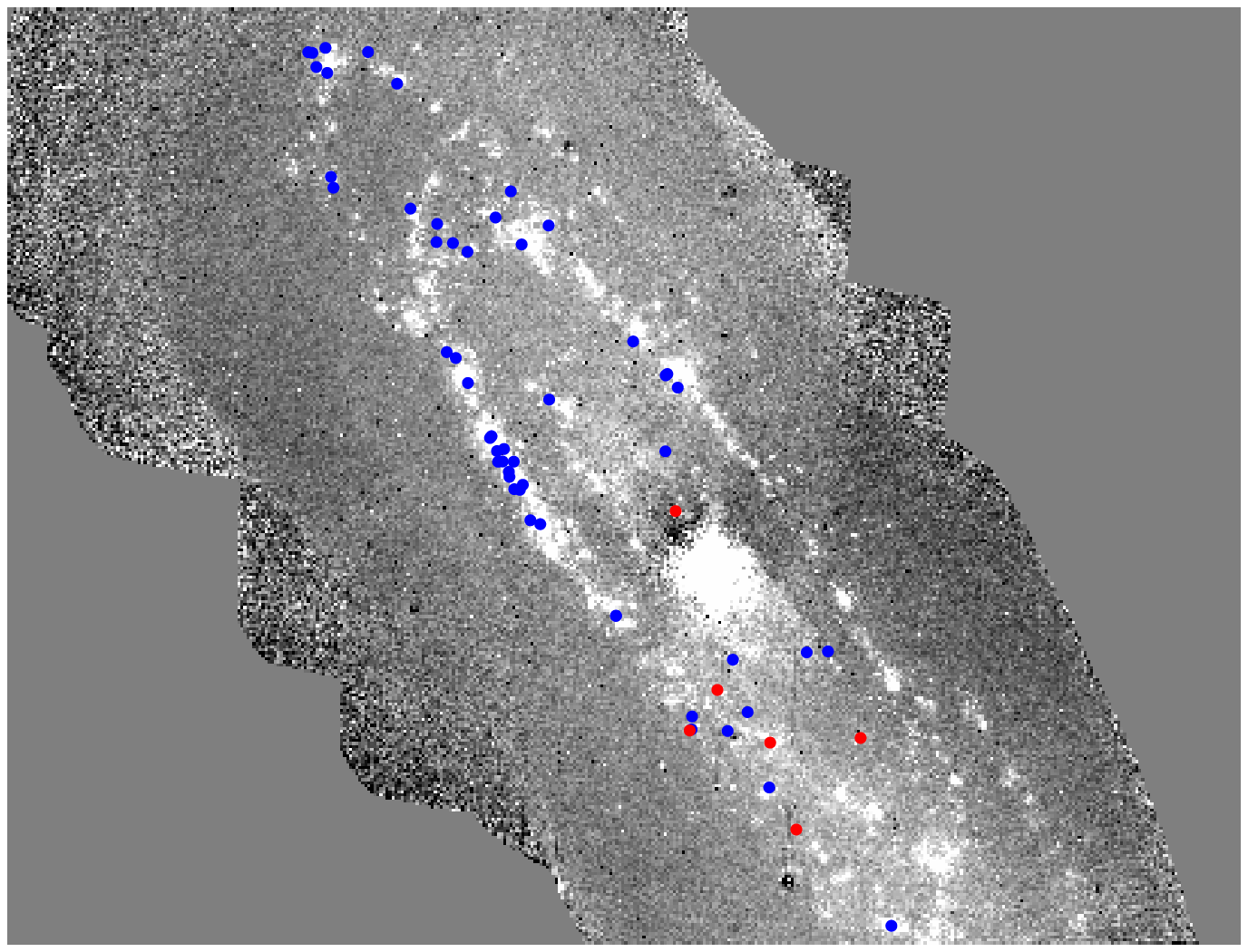}
    \caption{SN progenitor locations plotted on a star-subtracted
H$\alpha$ map of
M31. SNRs in red correspond to those with no young SF detected. We
consider these as possible Type Ia locations, or as the results
of runaway stars. We note that a large amount of our CCSN candidates
fall along the star forming arms of M31. This is clearly expected
(CCSN candidates will tend to be found in regions of recent SF),
but it also indicates the constraints imposed on our sample
by the locations of archival HST data (star-forming regions
are targeted significantly more than other areas).}
\label{spatial}
\end{figure}

In Table~\ref{tab_snr} 
we list the SNR for which our criteria is met, 
along with the corresponding HST fields. In cases where 
the regions in question were imaged multiple times, 
we selected whichever set of images had the greatest
number of stellar detections. In Fig.~\ref{spatial}
we chart the locations of the examined remnants on a star-subtracted
image of M31 in H$\alpha$. Those remnants colored red represent
probable Ia candidates, where no young SF was found. There is a clear
grouping of CCSN candidates along the star forming arms of M31.

We performed resolved stellar photometry
using the photometry pipeline developed 
for the ACS Nearby Galaxy Treasury program 
\citep{dalcanton2009}. This pipeline uses the DOLPHOT 
stellar photometry package \citep{dolphin2000} to fit 
the well-characterized ACS point spread function
to all of the point sources in the images. We then 
converted fluxes to Vega magnitudes using the 
standard zero-points and aperture corrections from the 
ACS handbook. We assess photometric errors 
and completeness using fake star tests. At least $10^{5}$ tests 
are performed by inserting fake stars of known color 
and magnitude into the data one at a time and blindly 
attempting to recover them with the same software.
Both fake star tests and photometry were performed
on the full HST fields.

\subsection{Calculation of Star Formation Histories}
CMD fitting is a powerful tool for measuring
star formation histories \citep{gallart2005,barker2007,williams2009c,
williams2009b,tolstoy2009}.
To estimate star formation histories of the 
regions surrounding the SNR, we used the software package 
MATCH \citep{dolphin2002}. MATCH works by creating many model 
CMDs based on theoretical isochrones for a variety of ages and 
metallicities. A linear combination of these model CMDs are then fit
to the observed CMD. We use the models of
\citet{marigo2008} and \citet{girardi2010} both for CMD
fitting and estimation of \mzams. 

For each field, we selected stars from our photometry catalog
within a 15'' radius around the 
SNR coordinates, which equates to a physical size of $\sim$50 pc
at an assumed distance modulus of $m-M=24.47$~\citep{stanek1998}.
This is consistent with the spatial correlation
discussed in \S2.1.
We account for photometric errors through the use of fake star tests.
We selected fake stars in a region $\sim$2.5 times the radius of 
the real star annulus to ensure that at least a few thousand recovered fake stars
are included for each SNR. We defined lower magnitude limits as the 
point at which fake star completeness dipped below 50\%.

MATCH requires a variety of parameters to generate and fit CMDs.
We assumed a Salpeter IMF, $dN/dM = M^{-2.35}$ 
\citep{salpeter1955}, 
and a binary fraction of 0.35. 
\citet{gogarten2009} demonstrated 
that varying the IMF value from -2.0 to -2.7 or varying the 
binary fraction from 0.2 to 0.5 had no effect
on the epoch assigned to a recent burst. We also varied the IMF from -1.3 to -3.3 for select regions
and found no significant difference in relative fractions of SF for various epochs.
Although absolute amplitudes of SF did change, only relative amplitudes
have an effect on our method of determining 
ages. Note that the IMF and binary fraction are used by MATCH
purely for purposes of populating the models. We assume nothing
about either value with regards to a potential CCSN progenitor
system or the overall progenitor mass distribution.

\begin{figure}
  \centering
  \epsfig{file=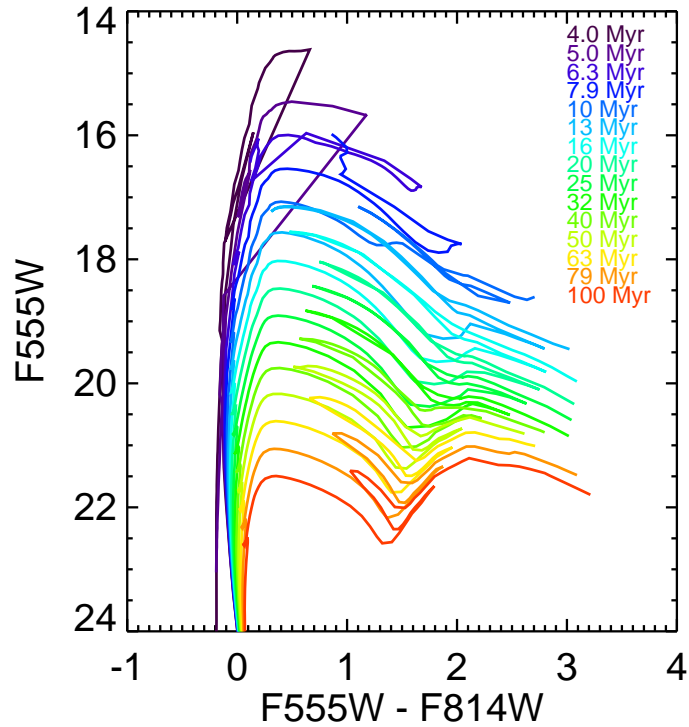,width=0.6\linewidth}
  \caption{\citet{marigo2008,girardi2010} isochrones plotted in
F555W against F555W - F814W for logarithmic
age bins of 6.60 through 8.00.  Redder isochrones correspond
to older ages. We adopt the 50 Myr isochrone as the maximum
age for a CCSN progenitor star, corresponding to a \mzams~of 7.3~\msun.
For clarity, we only plot every-other isochrone; our actual fitting procedure uses twice as
many age bins over the same range.}
  \label{isochrones}
\end{figure}

MATCH produces fits in logarithmic age bins. We adopted 71 age 
bins increasing in .05 increments from 6.60 (4 Myr) up to 10.10 (12.5 Gyr).
We are unable to fit
for ages younger than this due to the lack of isochrones at younger ages. MATCH
will interpret any SF from a population younger than the 6.60 to 6.65 (4 to 4.5 Myr) bin as being
included in this bin. Thus our youngest bin actually includes all SF from
present times back to 4.5 Myr.
As a result, we may only quote an upper age limit for any SF found in this youngest bin.
In Fig.~\ref{isochrones} we plot the isochrones for every-other age bin
from 6.60 through 8.00 in F555W vs. F555W-F814W. 

Fitting metallicity with MATCH is not viable given
the very weak dependence of the optical colors of the
upper main sequence on metallicity. We therefore constrained the
metallicity to a spread of $\sim$0.15 dex and to increase
with the lifetime of the galaxy, which consistently produced best-fit metallicities
of solar. This value is consistent with the known gas-phase metallicity
of M31 \citep{dennefeld1981,blair1982}.
Finally, we binned our CMDs in units of 0.3 in magnitude and 0.15 in color.
This binning accounted for the fact that some fields had few upper MS 
stars, allowing us to reduce the impact of this paucity on our 
fits. However, we still maintain a number of bins significantly 
larger than the number of free parameters used in the fitting.
We experimented with using finer binning and found no
change in our age determinations within our uncertainties.

\subsection{Treatment of Reddening}
Recent SFHs are very sensitive to the treatment of reddening. 
Our most simplistic treatment of reddening is to assume
that all reddening is due to a galactic foreground, search over
a specified range of reddening values for a best fit, and apply
this value to the CMD as a whole. However, we frequently found
significant differential reddening ($dA_{v}$) across our small SNR-centered regions,
which manifests as an increased width to the MS and red clump.
Star forming regions will typically have extensive
amounts of dust and gas, often not uniformly distributed, and as such
there is no reason to believe that a single reddening value would 
correctly describe the region as a whole.
MATCH by default allows for 0.5 mag of differential
reddening for populations below 40 Myr, after which it falls
linearly down to 0.0 mag at 100 Myr. In addition, the user
has the ability to add additional full-field differential reddening
across the CMD. We almost always found that
the default treatment of reddening was inadequate to account
for all the reddening present in the regions.

\begin{figure}
  \epsfig{file=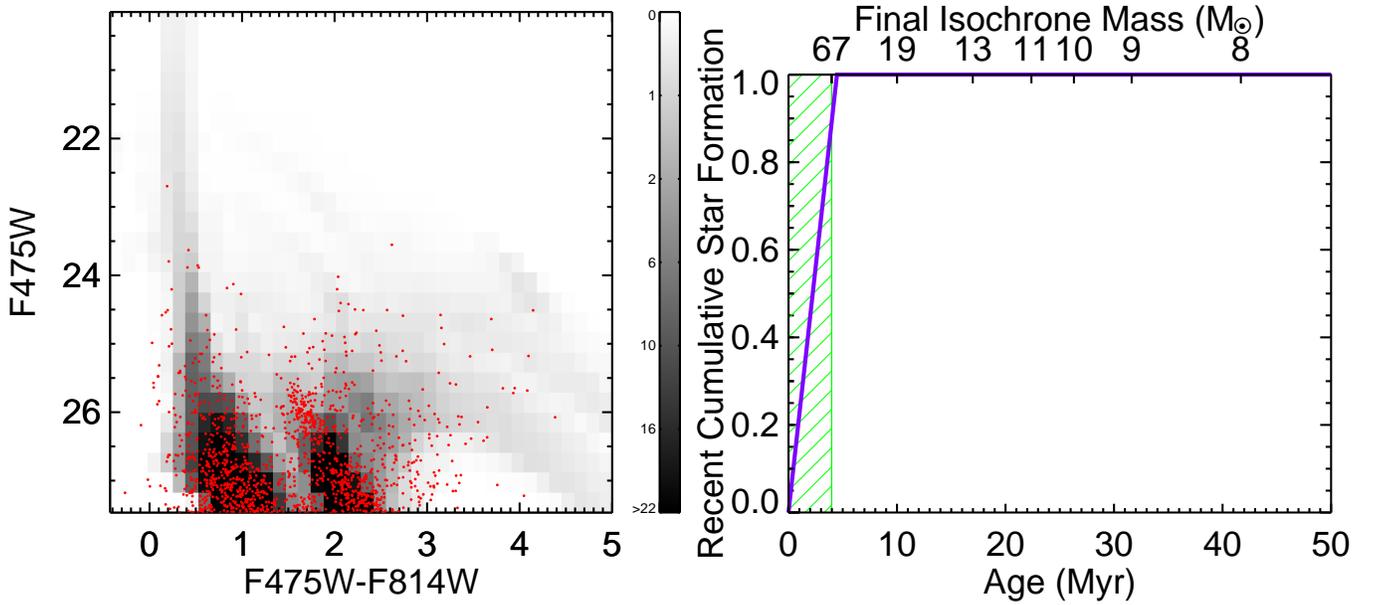,width=\linewidth}
  \caption{Left Panel: Observed CMD for the region around BW-102,
F814W plotted against F475W-F814W. In the background we plot
the best-fit model CMD generated by MATCH in greyscale, assuming no
additional differential
reddening. The model expects a much brighter MS than observed at $F475W<22.5$ and
poorly models the stars at $F475W\sim25.5$, $F475W-F814W\sim1.75$
when differential reddening is not accounted for.
Right panel: Cumulative recent star formation history calculated
for the corresponding CMD.  This SFH was from a blind run using 
MATCH's default values, with 0.5 magnitudes of differential reddening
applied only to the youngest ages. Based on this SFH, we may only
definite an upper limit on the age of 4.5 Myr or less, corresponding
to a mass of greater than 52 \msun. The CMD displays no
bright, blue upper MS stars, causing us to be suspicious of
the extremely young progenitor star. In addition, the extended
red clump indicates that differential reddening is not unique to
the young stellar population.}
\label{bw102_nodav}
\end{figure}

As an example of how ignoring differential reddening can lead to
an erroneous result, we consider the SNR BW-102.  The photometry
is quite deep, extending to magnitudes of $F475W=27.6$ and
$F814W=26.8$. In the left panel of 
Fig.~\ref{bw102_nodav} we plot the observed CMD of the region,
as well as the best-fit model generated by MATCH in the background in
greyscale. In the right panel, we plot
the cumulative SFH as a fraction of total SFH in the past 50 Myr, 
assuming a fixed differential reddening of $dA_{v}=0.5$ mags affecting
only the young stars. We found
a best-fit reddening value of $A_{v}=0.975$. All the star formation
is concentrated in
the youngest time bin ($<$ 4.5 Myr), corresponding to
a lower limit \mzams~of 52~\msun. However, several qualities of the CMD would
cause us to be suspicious of this result.
The main sequence appears very dim, and appears to contain lack
the young, bright, blue stars that the model predicts at $F475W<22.5$.
We would not expect this given
a result of very extensive young star formation.
Looking beyond the main
sequence, we see that the red clump is extended over nearly 2 magnitudes
in F475W. Indeed, the model very poorly models the stars
at $F475W\sim25.5$, $F475W-F814W\sim1.75$Thus not only was the default value
of differential reddening insufficient to model the entire CMD, the
treatment of differential reddening as something unique to the young
stellar population is clearly not reflected in the reality of the
CMD. For regions such as this, it is necessary to apply differential
reddening across the CMD. We must then define the amount of
differential reddening to add.

To understand our technique for determining the amount of $dA_v$ to use,
we must expand on the differential
reddening model used by MATCH. MATCH applies differential reddening to the model CMDs
by applying a top-hat distribution along the reddening line. The low
end of the top-hat is defined by a foreground value of $A_{v}$ for which MATCH
fits. The width of the distribution
is then defined by the specified value provided by the user.
When applying additional differential reddening,
fit values tended to improve until we reached
unphysical values of $dA_{v}$
The only true constraint we may apply is that this minimum
value of $A_{v}$ must be at least equivalent to the Milky Way foreground
value. \citet{schlegel98} find a foreground reddening value
to M31 of $A_{v}=0.19$. While this value is likely to change slightly
over the angular size of M31, the changes will clearly be small compared
to the magnitude of differential reddening internal to M31. Our solution to
CMDs such as BW-102 is to increase
the width of the differential reddening distribution applied to all ages until
the best-fit value of $A_{v}$ MATCH finds reaches the foreground value. We then
use this differential reddening distribution to measure the SFH
and derive the age of the SN progenitor. 

\begin{figure}
  \centering
  \epsfig{file=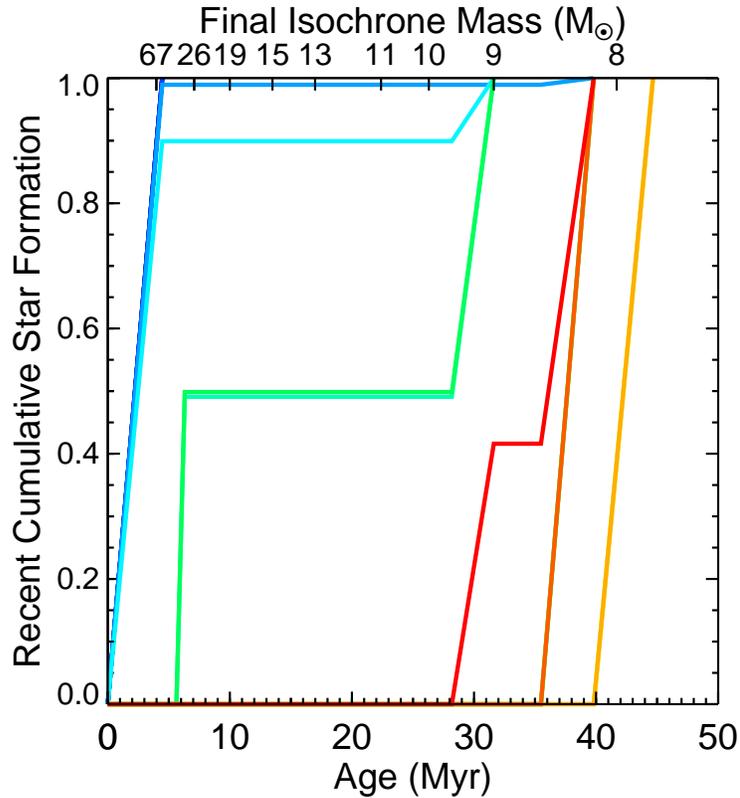,width=0.6\linewidth}
  \caption{Cumulative star formation history for BW-102.
Different lines correspond to different values of
field-wide differential reddening, with bluer lines corresponding
to $dA_v$=0.0 and redder lines corresponding to $dA_v$=1.5.  This is in addition
to the 0.5 magnitudes of $dA_v$ applied by default to the young stellar
population.  Note that some lines are not visible because they 
find the exact same age as other lines, and therefore show up
underneath other results. The red line was found to have the best fit
while still maintaining a value above that of foreground reddening.  It
corresponds to an age of 36 Myr and a mass of 8.5~\msun.}
  \label{bw102dav}
\end{figure}

\begin{figure}
  \epsfig{file=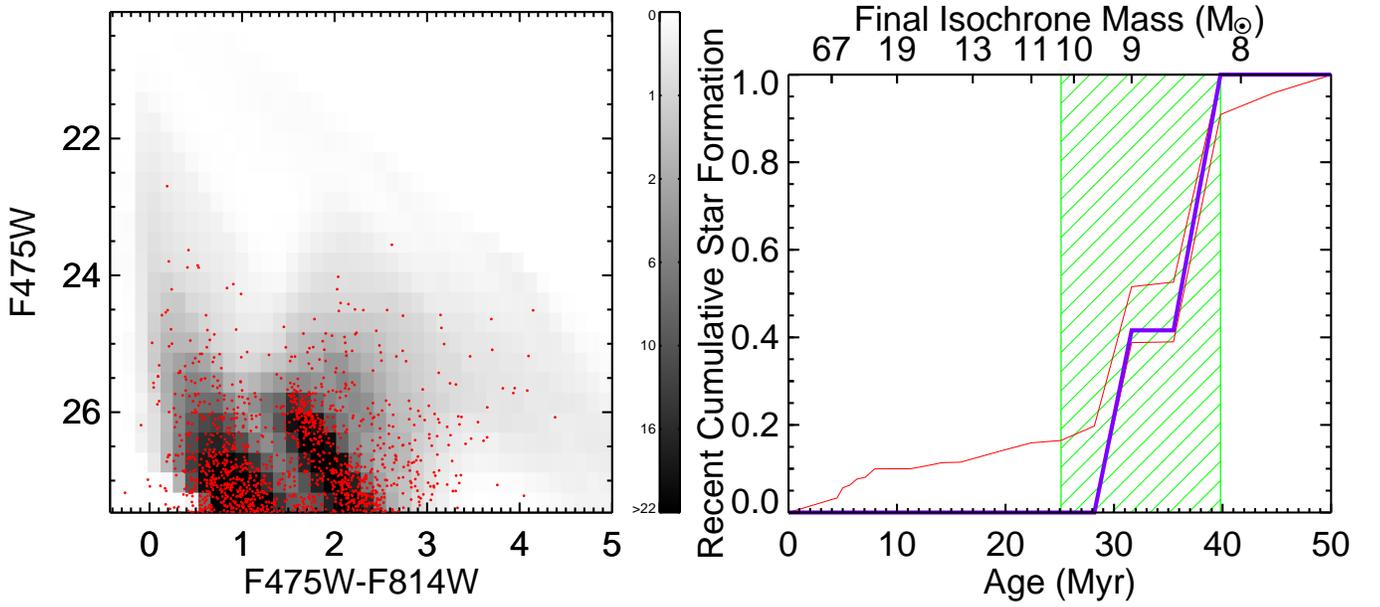,width=\linewidth}
  \caption{Left Panel: We plot the observed CMD of the region
surrounding BW-102, plotted as magnitude in F475W vs. (F475W-F814W).
We also plot the best-fit model created by MATCH in greyscale behind
the observed CMD. The scale is printed to the right of the CMD. This
model is found using the differential reddening procedure described in
\S2.4, leading us to use a value of $dA_v=0.5$. Right Panel:
Cumulative star formation over the most recent 50 Myr. We plot the
best fit (purple line) with high and low errors as calculated by the
Monte Carlo tests (red lines). The cross-hatched highlighted region
is corresponds to our 16\% to 84\% confidence interval, which we define
in \S2.3. We find
the data best fit by a burst of SF at $36^{+4}_{-11}$ Myr, corresponding
to $8.5^{+2.0}_{-0.5}$ \msun (highlighted region).}
\label{bw102}
\end{figure}

In Fig.~\ref{bw102dav} we plot the SFH results for a variety of
differential reddening values, increasing from $dA_{v}=0.0$ up to 
$dA_{v}=1.5$ magnitudes, in
increments of 0.1. Note that this differential reddening is in addition
to the default young $dA_{v}=0.5$, leading to up to 2.0 magnitudes of differential
reddening on the very youngest stellar populations.
We found that a value of $dA_{v}=1.5$ was the point at which the overall
reddening value dropped to the expected foreground value for BW-102. For this value of 
differential reddening, the SFH corresponded to an age of 36 Myr
and a mass of $\sim$9~\msun, a much more reasonable result given the observed
CMD. Note that the SFH reaches this result (within uncertainty) before the best-fit
$A_v$ value reaches foreground, suggesting that our result is not
highly contingent on the precise value of foreground reddening assumed.
In the left panel of Fig~\ref{bw102} we plot the observed
CMD with the best-fit model plotted in greyscale in
the background, created using the above $dA_v$ procedure.
In the right panel, we plot the cumulative SFH of the region
for the past 50 Myr, with error bars from the Monte Carlo analysis
described in \S2.5

We found essentially all fields suffered from 
differential reddening in excess of the default values used by MATCH.
When no extra differential reddening was included in the models,
many regions without obvious young star formation were best fit
by high overall reddening values and included
star formation in the youngest few time bin. 
The issue arises from faint main
sequence stars revealed by deep photometry. Without inclusion of differential
reddening, these stars are fit by the uniform foreground reddening value for the region,
which is typically a high value (e.g. $A_{v}>0.6$). Using this erroneously high value,
the stars are corrected to be brighter and
bluer than they actually are, leading MATCH to mistake them as evidence of
young star formation. By allowing for differential reddening, we can incorporate
the full range of different reddening values across the field.

For our final procedure, we increase the amount of full-CMD differential
reddening until we find the best fit $A_{v}$ reaches the foreground value,
$A_{V}=0.19$. We then adopt this value of differential reddening for subsequent
analysis of that particular SNR.

\subsection{Assessing Uncertainties}

\subsubsection{Multiple Coeval Populations}
Our ultimate goal is to estimate the mass of the progenitor.
We do this by identifying the
burst of recent star formation from which the progenitor star most 
likely originated, allowing us to assign an age to each progenitor
star. Ideally, each SFH
would have a single isolated burst, making it easy to associate
a single age to the young stellar population.
In many cases, however, multiple bursts of varying rates and durations
appear in the young star formation history. As a result,
we examine the cumulative star formation history of each
CMD as a fraction of the total recent star formation. When examined
as such, the fraction of star formation in each age bin corresponds
to the probability that the progenitor will be of that age. Because
we are only interested in recent SF, we examine only the most recent
50 Myr of SF (see \S4.1 for an explanation of this limit). 

\subsubsection{Fitting Errors}
Our results suffer from both 
random uncertainties due to the sampling of the CMD and systematic 
errors due to differences between the theoretical isochrones and 
observations. For example, if the models were consistently redder 
than the observed stars, we would generally find younger ages when
these models are applied to real data.
The random errors are generally highly dependent on the number of
upper main sequence stars in the field, with higher numbers 
providing a tighter constraint on the data (although more upper main
sequence stars may also simply indicate a younger population).
See \citet{gogarten2009} for further discussion of this point.

To analyze our uncertainties, we performed a series of
Monte Carlo (MC) realizations of the data (see also \citet{dolphin2002,weisz2011}).
In each MC run, we
re-sampled the observed CMD to account for the Poisson errors
of the stars. In addition, to estimate how systematic
model differences can impact the results, we add in random
shifts to the models for each run. We used
random shifts of $\sigma=0.02$ in temperature and $\sigma=0.17$
in bolometric luminosity.
These systematic offsets are much larger than uncertainty
in the distance modulus, thereby incorporating both effects
into our overall uncertainty analysis.

Uncertainty in the age of the
recent star formation  manifests itself in two distinct ways.
The first is uncertainty measured by the MC analysis performed
above, while the second is due to the width of the intrinsic 
spread of the burst across multiple age bins.
We estimate the latter of these as the difference
between the median age and the ages
where 16\% and 84\% of star formation has occurred. We estimate
the former of these as the RMS difference between the
median age of the best fit and
the median ages of SF from the MC
tests. We then add these differences
in quadrature to assign a confidence interval to each result. Thus
for each progenitor we determine a median age, as well as (potentially
asymmetric) uncertainties about this median. Finally, we make the
assumption that we are unable to determine the age of SF to a greater
precision than the age bins in which we have measured this SF. We
thus round the age range to the age of the next isochrone
out from the median. We perform this step for
both younger and older star formation, always rounding away from
the median. The ages of these isochrones determine
our final uncertainties.

In order to use this method of analysis, we must define
the maximum age of star formation which may
produce core-collapse progenitors. As explained in \S4.1,
we use our distribution to estimate a minimum mass for core-collapse
between 7.8 and 7.0~\msun. As a result,
we adopt 50 Myr as the maximum age of a core-collapse progenitor
and only perform the above mass estimation analysis over the
most recent 50 Myr of SF. 

\subsection{Converting SFH to Progenitor Mass}
The \citet{marigo2008} and \citet{girardi2010} models specify a maximum mass
for an isochrone at a given age. More massive stars will have already
died off. Thus the isochrones we have identified as bounding
our confidence interval can be linked directly to values
for \mzams. Our value for the median progenitor mass comes
from interpolating the final isochrone masses between
isochrones to the median age value. This is necessary
because our median age won't line up exactly
with a defined isochrone.

\begin{figure}
  \centering
  \epsfig{file=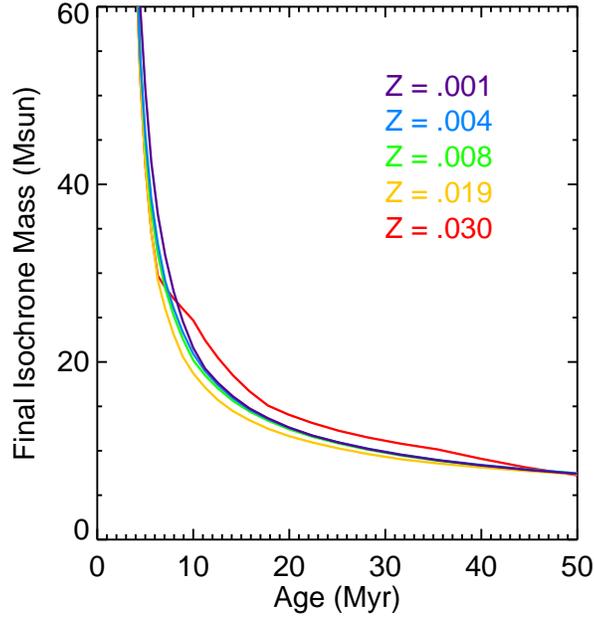,width=0.5\linewidth}
  \caption{\citet{marigo2008,girardi2010} final isochrone masses plotted
against isochrone age for metallicities of Z=0.004, Z=0.008
Z=0.019 (solar), and Z=0.030. Based on the known gas-phase metallicity
of M31 and our own best-fit metallicity values, we adopt solar metallicities
for all age-to-mass conversions in this paper.}
  \label{massplot}
\end{figure}

In Fig.~\ref{massplot} we plot the \citet{marigo2008} and \citet{girardi2010} 
isochrones for final isochrone mass vs.\ age for metallicities of 
Z = 0.004, 0.008, 0.019, and 0.030. The models produce very similar 
age to mass conversions regardless
of the assumed metallicity. The vast majority of our regions produce best
fit metallicities of approximately solar. As a result, we adopt the
Z = 0.019 isochrone for mass determinations. 
Note that masses change very
quickly for younger populations, while masses for older populations
change at a much slower rate. This means that our results for less
massive progenitors will naturally be more precise.

Note that we have neglected systematic uncertainties in the age-to-mass
conversion process for our individual progenitor results. This leads to
very small error bars on some progenitor masses, especially those at older
ages, where mass doesn't change significantly as a function of age.
Our mass distributions are not sensitive to this model-dependent systematic
uncertainty, as shown by similarity between the median progenitor
distribution and the probability distribution (see \S4.2).
However, uncertainties are almost certainly underestimated for some progenitors,
especially at the low-mass end.  To estimate the magnitude of this systematic
uncertainty, we compared the solar metallicity isochrones of \citet{pietrinferni2004}
to the \citet{marigo2008,girardi2010} isochrones used for our analysis.
We found that for a $\sim$50 Myr lifetime star, the \citet{pietrinferni2004} isochrones find
a maximum mass of  6.7~\msun, compared to the 7.3~\msun predicted for the
\citet{marigo2008,girardi2010} isochrones.  At a lifetime of $\sim$22 Myr,
the youngest stars tested by \citet{pietrinferni2004}, the maximum masses for the different
isochrone sets are 9.9~\msun and 10.9~\msun respectively.  These suggest
systematic uncertainties of around 0.5-1.0~\msun for the age-mass conversion
process. Again, we do not include these systematic uncertainties in the
reported values in Table 2; the reader should keep this in mind when
interpreting the results of any individual progenitor star in our study.

\section{RESULTS}
In Table~\ref{tab_results} we list results for all 59 SNR analyzed. We tabulate
the designation, mass, age, number of main sequence stars (defined as
$F475W-F814W\le0.4$, $F555W-F814W\le0.4$, and $F606W-F814W\le0.3$),
total stellar mass, and additional full-field $dA_{v}$ applied.
In general the number of MS stars may be taken as an indication
of confidence in the answer, but older populations will also generally have fewer MS stars.
As a result, this mapping is very approximate. The total stellar mass listed is highly
dependent on the precise value selected for the IMF,
and as such should not be taken as an exact measurement. It is intended
only to compare relative amplitudes of star formation between various SNR regions.
We observe six fields that have no significant recent SF, and identify these as likely either
Type Ia SNR or SNR resulting from runaway progenitor stars. They are not included
in analysis of the distribution of recovered progenitor masses.

\begin{figure}
  \centering
\epsfig{file=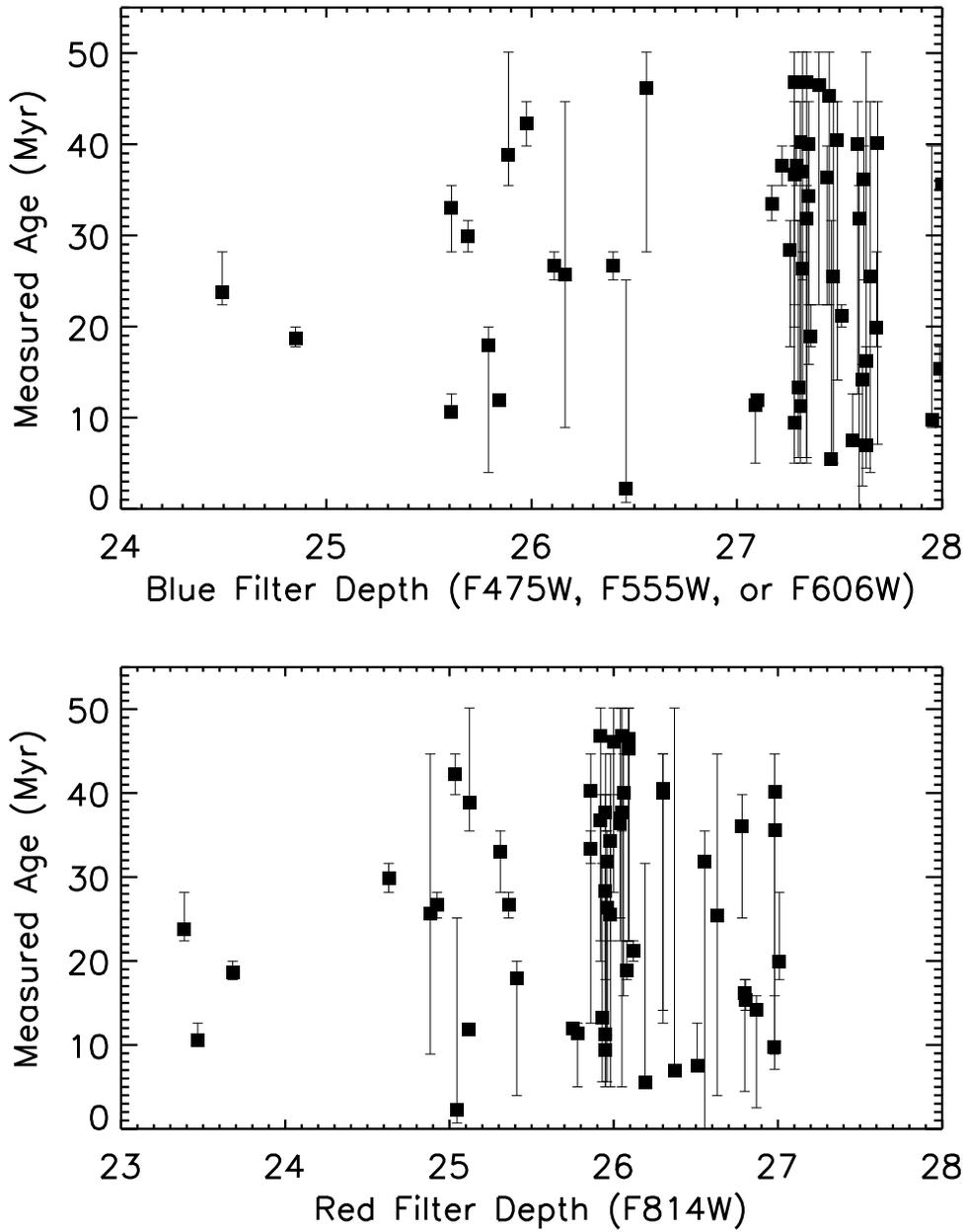,height=7in}
\caption{Top Panel: We plot the recovered ages for the progenitor
stars against the 50\% completeness limit for the redder broadband
filter used. Bottom Panel: Same as top, but for the bluer
filter used. The filter selection is somewhat arbitrary, since
the bluer filter may be F475W, F555W, or F606W depending
on the dataset. The redder
filter will always be F814W. The data points do not include
results for which no young SF was found.
Following our depth cut, we find no significant correlation
between depth in either filter and the measured age.}
\label{depthcompare}
\end{figure}

To examine any underlying biases in our data selection, we plot
the recovered
ages against the 50\% completeness magnitudes in Fig.~\ref{depthcompare}.
The top panel displays the bluer filter for each particular CMD,
while the bottom is for the redder (F814W in all cases).
Following application of our depth requirement, we find no evidence
of any correlation between recovered age and
depth of data.

\begin{figure}
  \centering
\epsfig{file=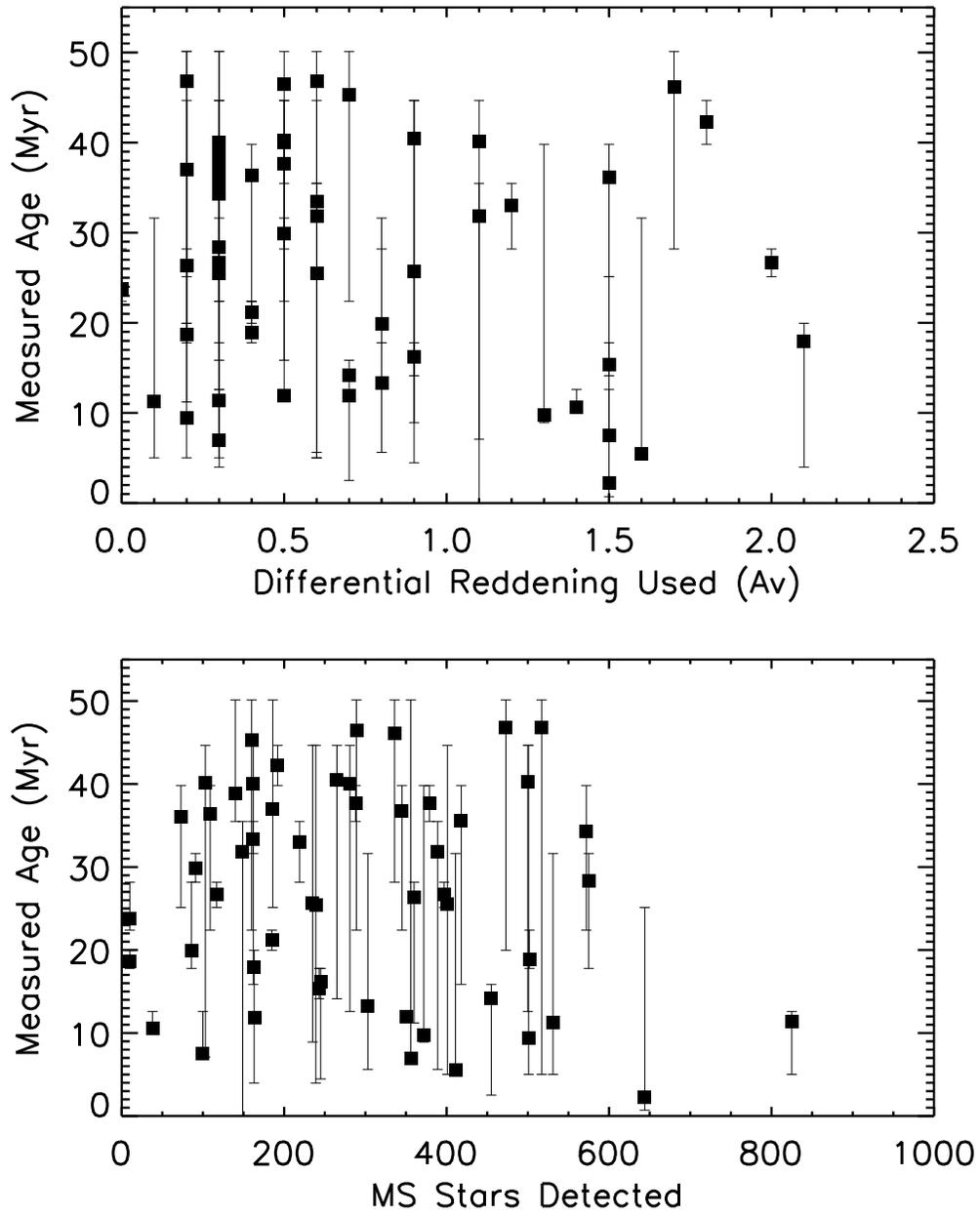,height=7in}
\caption{Top panel: We plot the recovered ages for the progenitor
stars against the number of MS stars detected in the region around
the SNR. Bottom panel: We plot the recovered
ages against the value of $dA_{v}$ used for ACS data only. These
values include only the additional user-added $dA_{v}$,
not the 0.5 mag of $dA_{v}$ automatically included for young populations.
The data points do not include results for which no young SF was found.
We find no significant correlation between age
and either $dA_{v}$ or number of stars.}
\label{mscompare}
\end{figure}

The top panel of Fig.~\ref{mscompare} plots the recovered ages
as a function of the number of MS stars detected.
The bottom panel
plots recovered ages as a function of $dA_{v}$ used.
Note that these values only include full-field $dA_{v}$, not
the 0.5 mag of $dA_{v}$ included for all young populations.
No correlation is observed in either
comparison, which suggests
that our treatment of differential reddening from \S2.4
does not bias us towards a specific age.

While an in-depth discussion of all SNR is prohibitively long, it is
worth exploring a representative sample of the various cases
observed. In general, fields with fewer main sequence
stars will tend to have larger uncertainties, and we list our results
in Table ~\ref{tab_results} in order of the number of these stars.
However, this generalization is not absolute,
and this ordering should be taken as approximate. Finally, some of the regions display
no young star formation, which we list in a separate category.  
For illustrative purposes, we subjectively identify
four classes of results. Below we examine four representative SNR for
each case.

\subsection{K376: Obvious Young Star Formation}
\begin{figure}
  \epsfig{file=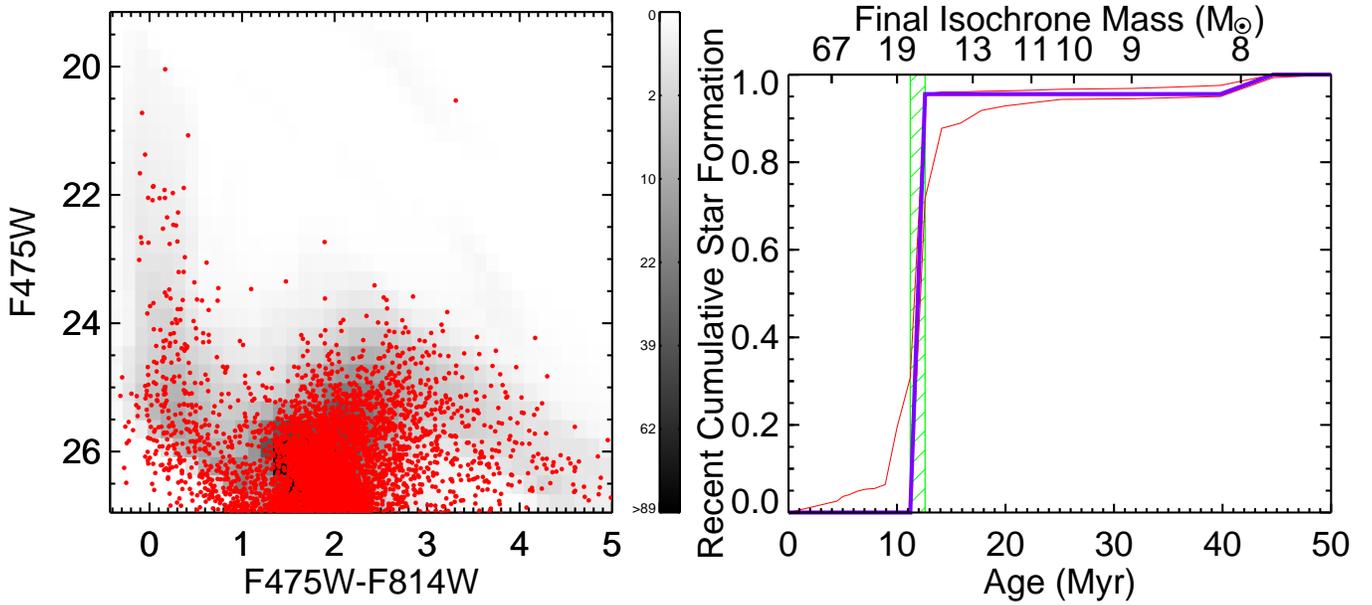,width=\linewidth}
  \caption{Left Panel: In red we plot the observed CMD of the region
surrounding K376, plotted as magnitude in F814W vs. (F475W-F814W).
We also plot the best-fit model created by MATCH in greyscale behind
the observed CMD. The scale is printed to the right of the CMD. A prominent
upper main sequence is clearly visible in the CMD, indicating
a young stellar population and a more massive progenitor. Right Panel:
Cumulative star formation over the most recent 50 Myr. We plot the
best fit (purple line) with high and low errors as calculated by the
Monte Carlo tests (red lines). The cross-hatched highlighted region
is corresponds to our 16\% to 84\% confidence interval, which we define
in \S2.5. We find
the data best fit by a single burst of SF at 12$\pm$1 Myr, corresponding
to 16$\pm$1 \msun (highlighted region).}
\label{k376}
\end{figure}

In the left panel of Fig.~\ref{k376}  we display the color-magnitude diagram for the region
around K376, plotted in red. We also plot our best-fit model CMD in the background
in greyscale, with darker regions corresponding to a larger expected
stellar density. The data we used for K376 are from HST project
12055 (Brick 9, Field 14), and included ACS images in F475W and F814W. The photometry
is fairly deep, with 50\% completeness limits at F475W=27.1 and 
F814W=25.8. We find 6874 stars in the 50 pc region around the SNR, 350 of
which are MS stars, making K376 a very well populated CMD.

A qualitative inspection of the CMD indicates a bright, blue upper main
sequence, indicative of a significant young stellar population. The
CMD displays a great deal of differential reddening, as indicated
by a large spread in the red clump. We used a full-field value of $dA_v=0.7$,
in addition to $dA_{v}=0.5$ for young stars, to
model this CMD. MATCH predictably
finds that this CMD displays significant young star formation: we find the
field well fit by a single burst of star formation at 12$\pm$1 Myr, corresponding
to a mass estimate of 16$\pm$1 \msun\. The older burst at $\sim$40 Myr is of 
significantly lower prominence compared to the young burst. The result
is well-constrained due to the large number of stars in the field.

Similar SNR include 2-020, 2-024, K934, K947, 
BW-60, BW-69, BW-74.

\subsection{K180: Well Defined Older Populations}
\begin{figure}
  \epsfig{file=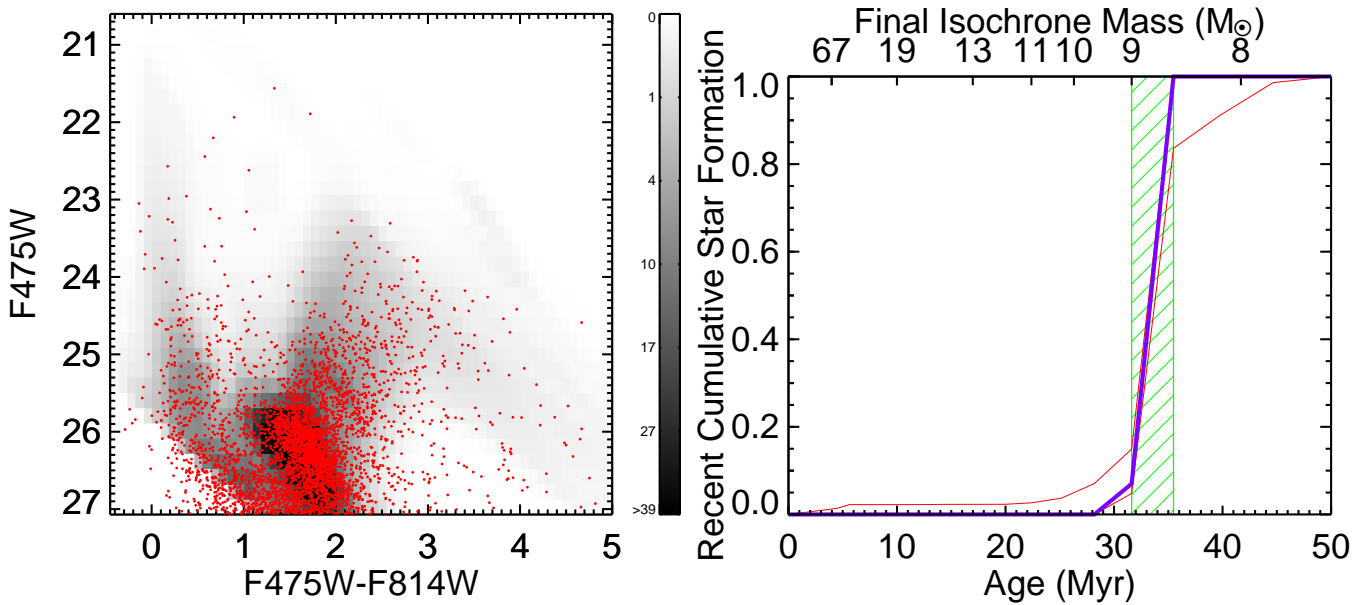,width=\linewidth}
  \caption{Same as Fig.~\ref{k376}, but for K180.
Left Panel: In red we plot the observed CMD of the region
surrounding K180, plotted as magnitude in F814W vs. (F475W-F814W).
We also plot the best-fit model created by MATCH in greyscale behind
the observed CMD. The scale is printed to the right of the CMD. 
The main sequence is dimmer than that for K376, indicating
an older population and thus a less massive progenitor. Right Panel:
Cumulative star formation over the most recent 50 Myr. We plot the
best fit (purple line) with high and low errors as calculated by the
Monte Carlo tests (red lines). The cross-hatched highlighted region
is corresponds to our 16\% to 84\% confidence interval, which we define
in \S2.5. We find
the data best fit by a single burst of SF at 33$\pm$2 Myr, corresponding
to 8.8$\pm$0.2 \msun (highlighted region).}
\label{k180}
\end{figure}

Fig.~\ref{k180} displays the CMD and SFH for the
region surrounding K180. The data for K180 are from 
HST project 12073 (Brick 2, Field 11), and includes ACS images in F475W and F814W.  The photometry
is quite deep, with 50\% completeness
limits of F475W=27.2 and F814W=25.9 respectively. We find 3950 stars 
in the CMD, 162 of which are MS stars. We used full-field value of $dA_v=0.6$,
in addition to $dA_{v}=0.5$ for young stars, to model this CMD.

The upper main sequence is not nearly as bright
or prominent as that of K376, and as such
we would predict an older population with a lower mass progenitor star.
MATCH agrees, finding the population is fit best by a single burst of star
formation at 33$\pm$2 Myr, corresponding to a mass of 8.8$\pm$0.2~\msun.
As above, because of the large number of stars in the field,
the mass is fairly well constrained.

Similar SNR include
1-006, 1-008, 1-009, 1-010, 2-025, 2-044, 2-046, 2-048,
2-050, K516, K526A, K574, K594, BW-18, BW-20, BW-31, BW-32,
BW-76, BW-81, BW-84, BW-86, BW-102, and BW-110.

\subsection{K891: SF Spread Over a Wide Range}
\begin{figure}
  \epsfig{file=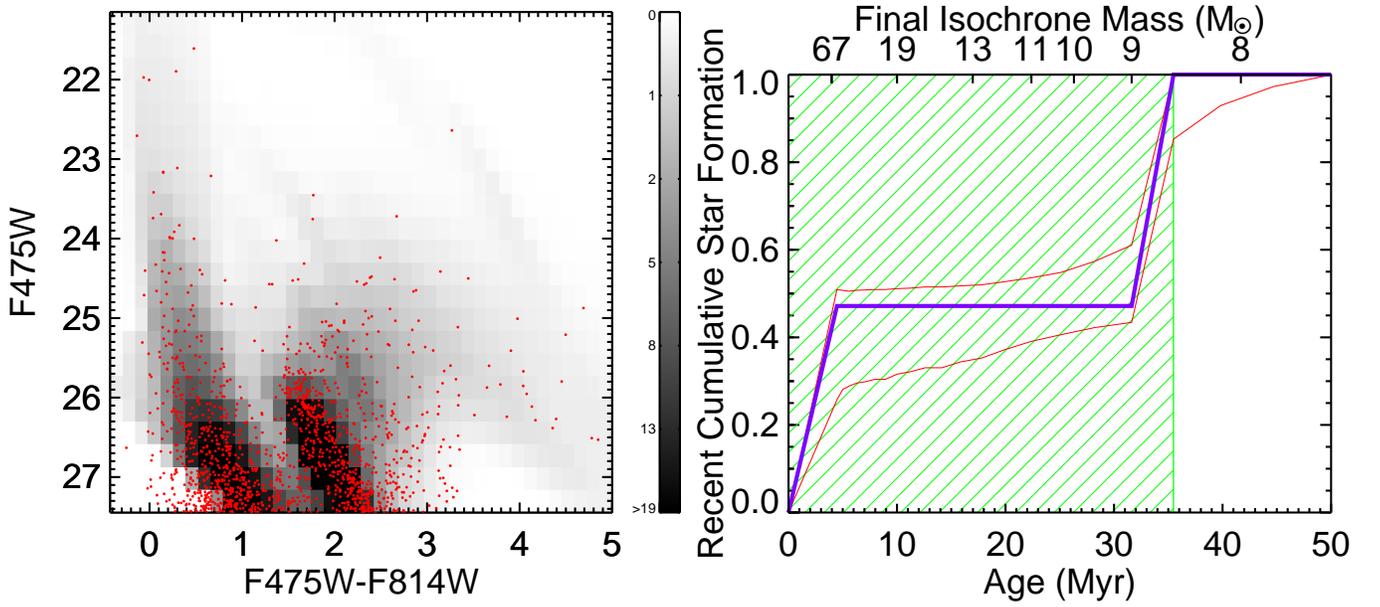,width=\linewidth}
  \caption{Same as Fig.~\ref{k376}, but for K891.
Left Panel: In red we plot the observed CMD of the region
surrounding K891, plotted as magnitude in F814W vs. (F475W-F814W).
We also plot the best-fit model created by MATCH in greyscale behind
the observed CMD. The scale is printed to the right of the CMD. Right Panel:
Cumulative star formation over the most recent 50 Myr. We plot the
best fit (purple line) with high and low errors as calculated by the
Monte Carlo tests (red lines). The cross-hatched highlighted region
is corresponds to our 16\% to 84\% confidence interval, which we define
in \S2.5. Unlike the previous examples, K891 seems to display
two star formation events of similar prominence. We are unable
to associate the progenitor star with either burst, and as such
the only constraint we may offer is to say that the progenitor is
younger than 36 Myr, corresponding to a mass of $\ge9$~\msun.}
\label{k891}
\end{figure}

Fig.~\ref{k891} displays the CMD and SFH for the region surrounding
K891. The data for K891 are from 
HST project 12055, including ACS images in F475W and F814W.  The
photometry is again quite deep, with 50\% completeness limits of
F475W=27.6 and F814W=26.6 respectively.  The CMD has
1441 total stars, including 149 MS stars.
We used a full-field value of $dA_v=1.1$, in addition
to $dA_{v}=0.5$ for young stars, to model this CMD.

The CMD for K891 doesn't appear to comprise a single uniform population,
and indeed has stars spread out at a wide variety of ages. MATCH finds
two distinct star formation bursts: one in our youngest age bin at 
less than 4.4 Myr, and another in an older bin at $\sim$32 Myr. The
error bars also indicate that the relative prominence of these bursts
is such that we may not favor one burst over
the other. While we find a median mass of $\sim$9~\msun, the uncertainties
allow this result to range over all almost masses past the minimum mass
for core-collapse. K891 is an example of a SNR where we have reasonable
confidence in our answer, but this answer allows no constraint on the
parameters we ultimately wish to measure.

Similar SNR include 2-049, K446, K497, K525A, K527A, K856A,
K908, K956A, BW-11, BW-39, BW-44, BW-61, BW-65,
BW-66, BW-71, BW-77, BW-82, BW-89, BW-105, and BW-106.

\subsection{2-028: No Recent Star Formation}
\begin{figure}
  \epsfig{file=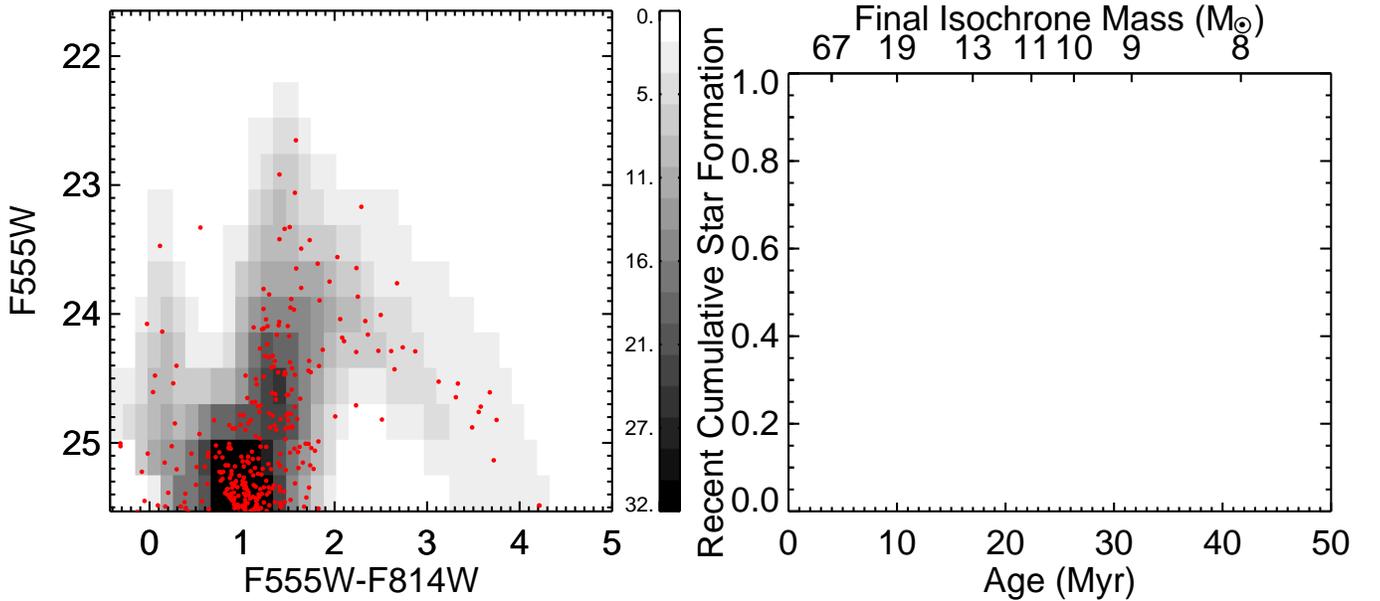,width=\linewidth}
  \caption{Same as Fig.~\ref{k376}, but for 2-028, with F555W
instead of F475W. Left Panel: In red we plot the observed CMD of the region
surrounding 2-028, plotted as magnitude in F814W vs. (F555W-F814W).
We also plot the best-fit model created by MATCH in greyscale behind
the observed CMD. The scale is printed to the right of the CMD. 
The main sequence is much dimmer than that for K376, indicating
an older population and thus a less massive progenitor. Right Panel:
Cumulative star formation over the most recent 50 Myr.
We find no SF within the past 50 Myr for 2-028. We classify
2-028 as either a possible Type Ia remnant, or as a possible
runaway star.}
\label{2-028}
\end{figure}

Fig.~\ref{2-028} displays the CMD and SFH for the region surrounding
2-028. The data for 2-028 is
also from HST project 10273, and includes ACS images in F555W and F814W.
The photometry has more shallow 50\% limits than the CMDs above, with
F555W=25.7 and F814W=25.4 respectively.  The CMD has only 374 stars, a smaller
number of detections than typical, including 53 MS stars.
We used a full-field value of $dA_v=0.5$,
in addition to $dA_{v}=0.5$ for young stars, to model this CMD.

The CMD clearly has a very sparse and dim main sequence, leading us to 
predict that MATCH would find little to no young star formation.
Indeed, MATCH finds the youngest star formation as occurring at
older than 90 Myr, well beyond the age of stellar populations
that we would expect to produce progenitors of core-collapse SNe.
We classify this SNR as the result of either a Type Ia SN
or a runaway star that has left the young stellar population
at its birth site. We do not use any results in this category
as for purposes of analyzing the mass distribution.   

Similar SNR include 2-016, 2-021, 2-026,
BW-19, and BW-36.

\section{DISCUSSION AND SUMMARY}
\subsection{Indications of a Minimum Mass}
Our method measures the prominence of a burst as a fraction 
of the total recent SF. This approach requires that we define 
the period of time that we consider as ``recent'' SF. Specifically,
we must identify the age range where star formation can produce
stars massive enough to result in CCSNe. 
Theoretical arguments and observational evidence point
to a minimum mass necessary for progenitors to undergo core-collapse
to a neutron star. Stars below this mass
are generally assumed to leave behind white
dwarf stars, producing no SN explosion. Thus one may constrain
this cross-over mass by either measuring the maximum mass
from which a star may create a white dwarf, or the minimum
mass necessary for a star to explode. 
Measurements of white dwarfs 
have defined a lower limit on this minimum mass of 
6.3-7.1~\msun \citep{williams2009}, corresponding to an age of
between $\sim$55 and 63 Myr, and direct progenitor mass
measurements has have converged on a value
of 8$\pm$1~\msun \citep{smartt2009,botticella2012},
corresponding to an age of between $\sim$33 and 55 Myr.   

In our observed progenitor mass distribution, we would
expect to see the following behavior: above the minimum
mass for core-collapse, we expect the distribution of progenitor
masses to follow the IMF, assuming that the recent SFR is approximately
constant. Below the minimum mass, the inferred progenitor mass should
have no physical connection to the CCSNe process and should reflect
random sampling of the SFR at $>$50 Myr, producing an essentially
flat distribution in inferred progenitor mass.

\begin{figure}
  \centering
  \epsfig{file=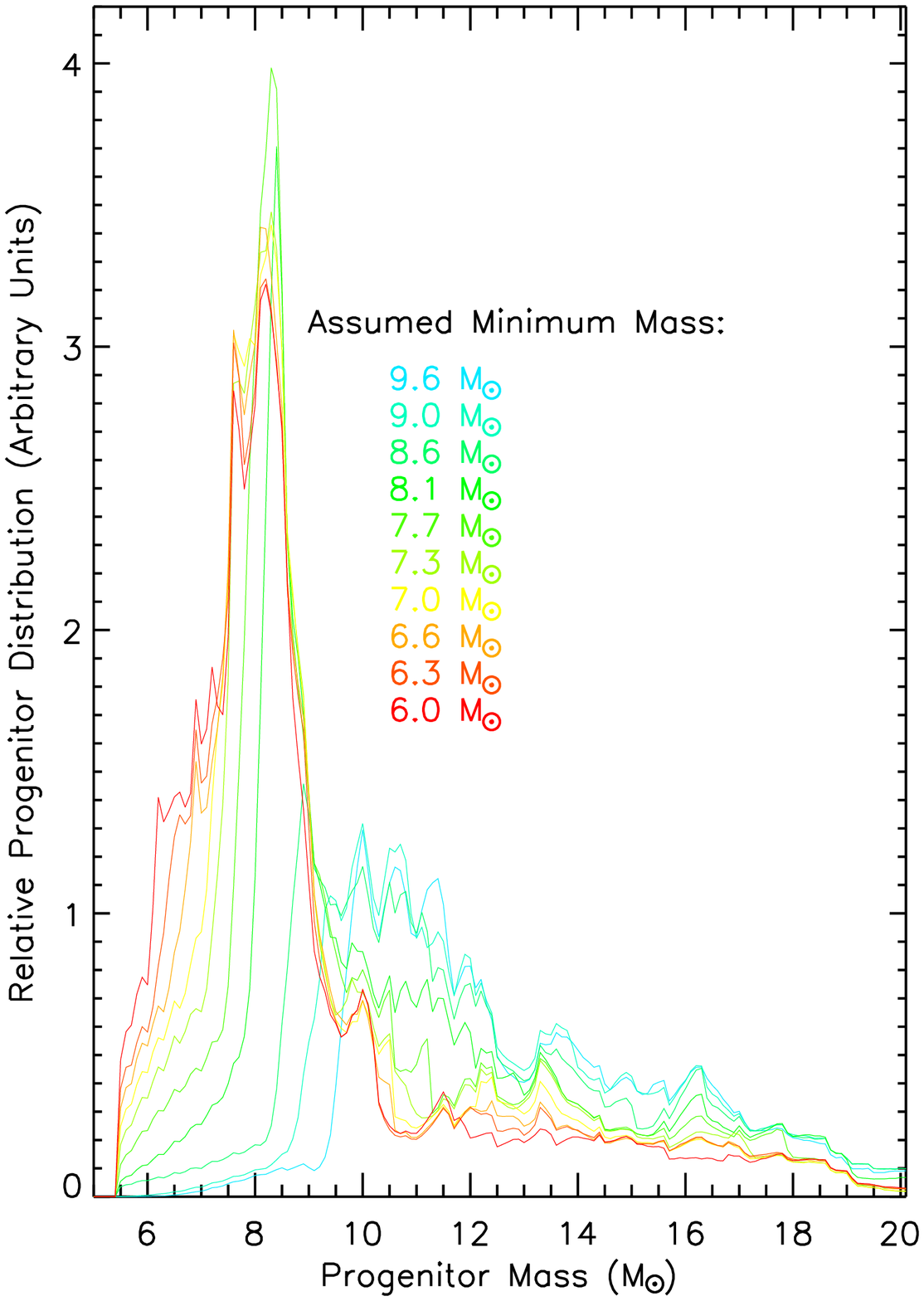,height=4in}
  \epsfig{file=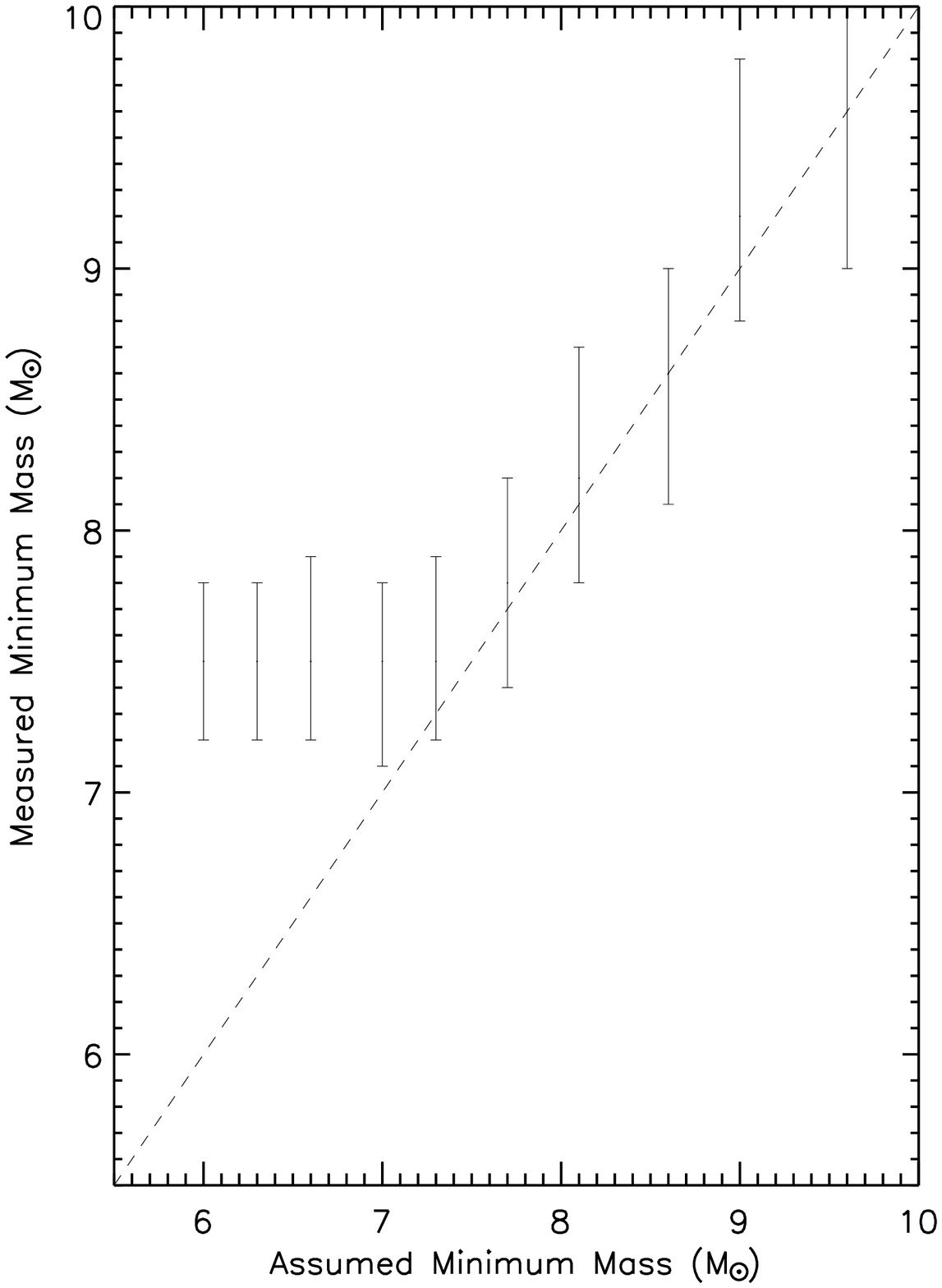,height=4in}
  \caption{Left Panel: Plot of distribution of progenitor masses
assuming different minimum mass for a star to undergo
core collapse. We plot distributions for 6.0, 6.3, 6.6,
7.0, 7.3, 7.7, 8.1, 8.6, 9.0, and 9.6~\msun. We note the existence
of a large peak of masses located at $\sim$8~\msun, which we argue
represents the minimum core-collapse mass. We also note that
this peak grows in size as we lower the minimum mass down to
7.3~\msun, where it is constant. This suggests that we include
the full number of progenitors when we include all star formation
down to 7.3~\msun, equivalent to the past 50 Myr of star formation.
Right Panel: On the horizontal axis, we plot the minimum
mass that we consider for purposes of determining recent
star formation. On the vertical axis, we plot the corresponding mass
for the largest value in the derivative of the progenitor
mass distribution with respect to mass. We argue this spike represents
the minimum mass for a star to undergo core-collapse. We find that
above 7.3~\msun, this value follows the assumed minimum mass. For 7.3~\msun and
below, this value remains at $\sim$7.5~\msun. We assume that the precision
with which we may define a value is $\pm$ one age bin, and that the error
bars on this minimum mass correspond to the final isochrone mass for these
bins. Using this analysis, we find a range of values for the minimum
mass from 7.0 to 7.8 \msun.}
  \label{minmass}
\end{figure}

In the left panel of Fig.~\ref{minmass}, we plot the distribution of progenitor masses
for a variety of assumed minimum masses. We vary the minimum mass from
9.6~\msun to 6.0~\msun (our chosen masses are mapped using the~\mzams\
from \citet{marigo2008} and \citet{girardi2010} isochrones). We find
that for assumed minimum mass greater than 8.1~\msun, the distribution increases until the assumed minimum
mass is reached. For assumed minimum mass values below 8.1~\msun, there is
a peak between 7.5 and 8.5~\msun, below which the number of progenitors
drops, suggesting a minimum mass in this range.

To find the actual minimum mass, we lower the assumed minimum mass
until the measured minimum mass no longer reflects this assumed value.
We note the amplitude of the peak in the distribution
is greatest for an assumed minimum of 7.3~\msun. To quantify the
location of this peak, we calculate the derivative of the number of progenitors
as a function of mass. We assume that the maximum value of this derivative
occurs at the minimum mass a star undergoes core-collapse, as this value identifies
the beginning of the peak. In the right 
panel of Fig.~\ref{minmass}
we plot the location of this maximum against our assumed minimum mass. We
note that above an assumed minimum mass of 7.3~\msun, the peak value traces the assumed minimum
mass. For 7.3~\msun~and below, however, the peak value is always around
$\sim$7.5~\msun. If we assume our uncertainties are at least the width of the
mass bins, the we find a minimum mass for core-collapse between 7.0 and 7.8~\msun.
This range is consistent with the observational measurements of
\citet{smartt2009} and \citet{botticella2012}. We therefore have adopted
the 44.7 to 50 Myr (7.7 to 7.3~\msun) bin as the oldest included for
our final results.

\subsection{Progenitor Mass Distribution}
\begin{figure}
  \centering
\plottwo{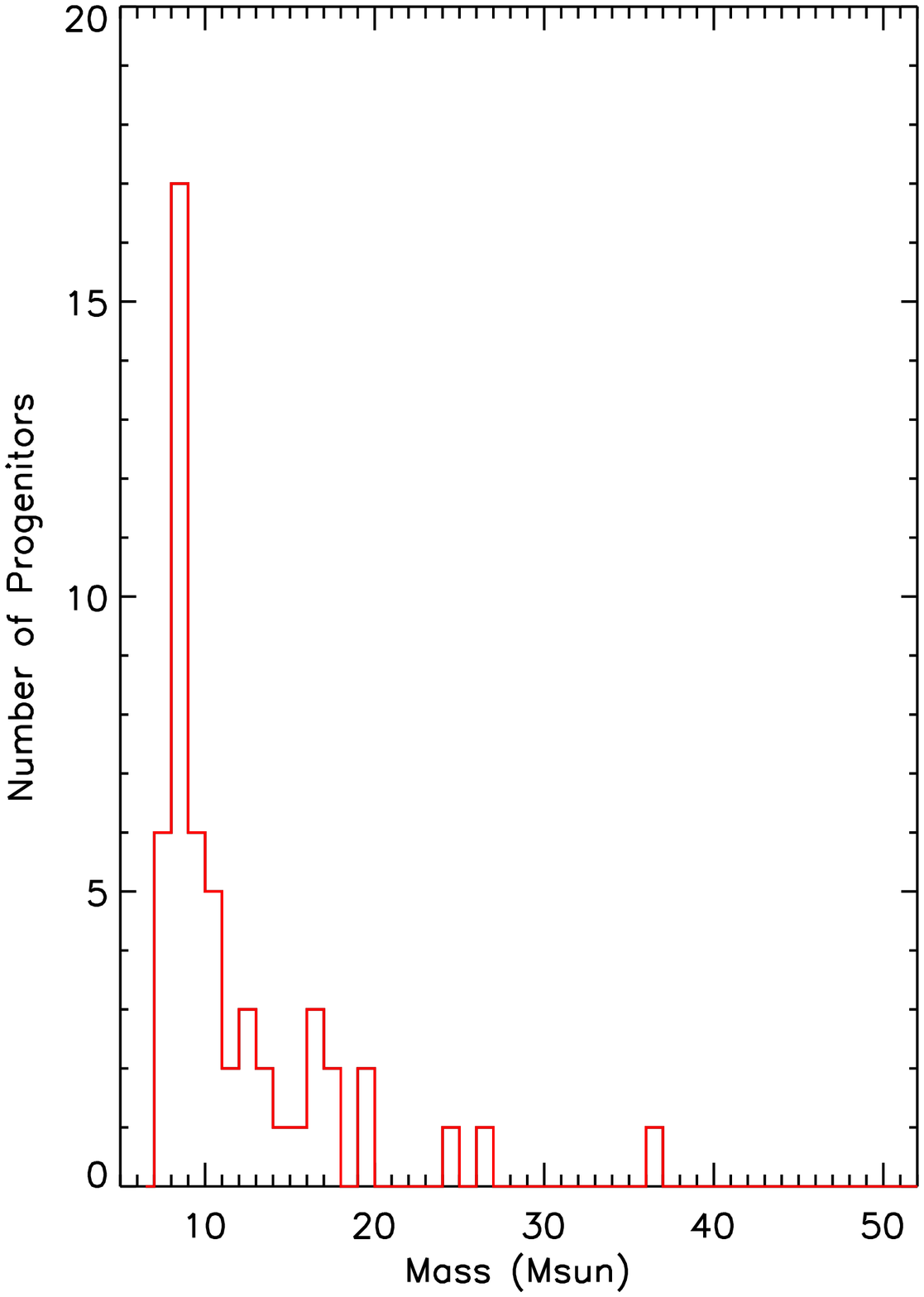}{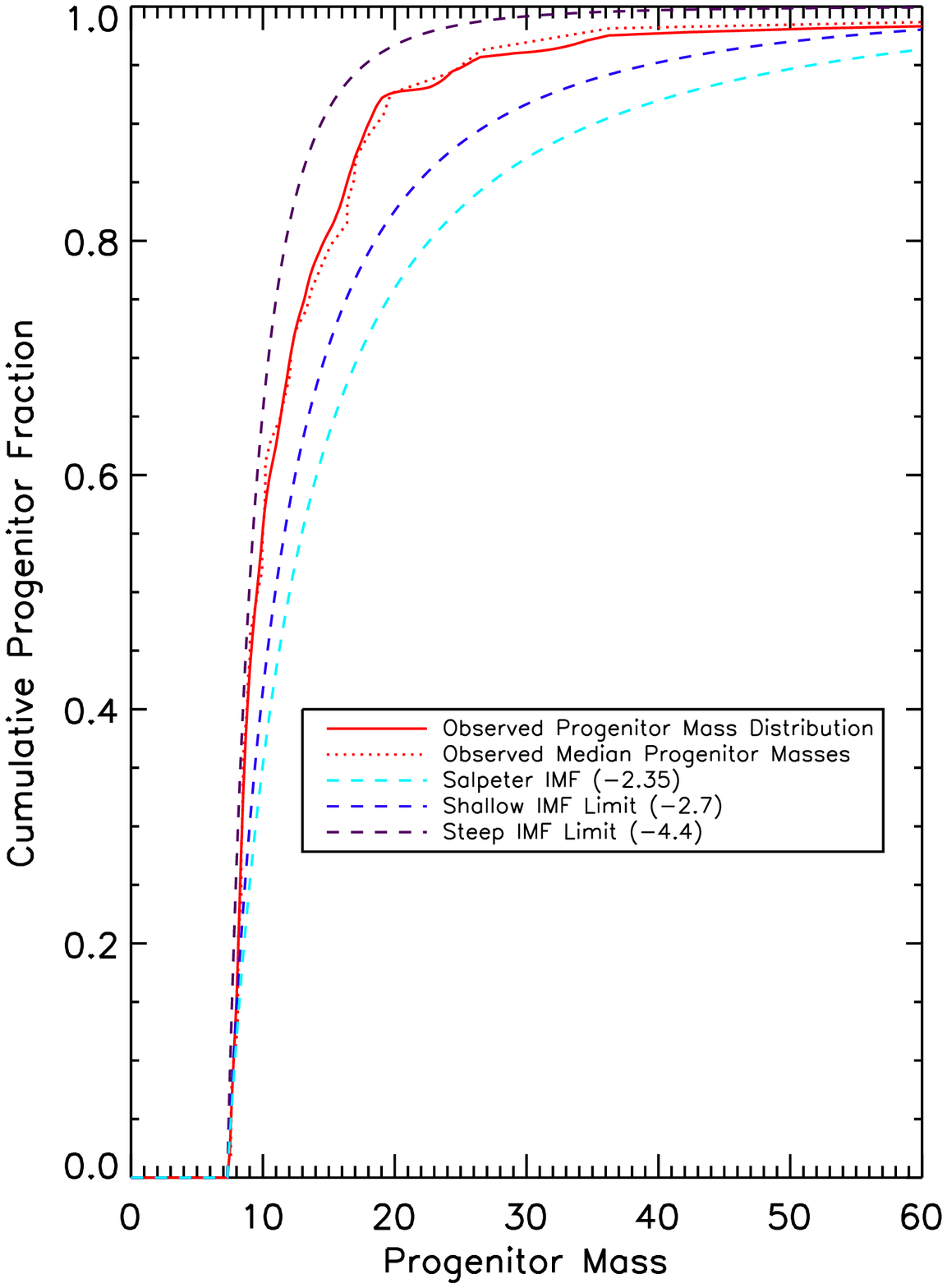}
  \caption{Left Panel: Histogram of median progenitor masses
below 52 \msun.
Right Panel: Cumulative fraction of progenitor mass distribution. We
overplot a reference Salpeter IMF. Using a KS-test, we find the
cumulative mass distribution consistent with a power-law IMF
of the form $dN/dM \propto M^{\alpha}$ with $-2.7\ge\alpha\ge-4.4$.
We plot these two slopes, as well as a Salpeter
IMF ($dN/dM \propto M^{-2.35}$).
While the distribution of masses greater than 60 \msun~is not
shown, the fraction greater than 60 \msun~is given by the value for
the cumulative fraction at 60 \msun.}
  \label{diffhist}
\end{figure}

In the left panel of  Fig.~\ref{diffhist} we plot the 
histogram of median progenitor masses, restricted to
7~\msun~and above.
In the right panel of Fig.~\ref{diffhist} we plot the
cumulative fraction of  progenitor masses. Unless otherwise
noted, we assume a maximum mass of 120~\msun, although
the choice of this value at high masses is essentially irrelevant
to the overall distributions given the rarity of extremely massive stars.

Qualitatively, the observed distribution
shows a lack of the most massive stars when compared to a
Salpeter IMF ($dN/dM \propto M^{\alpha}$, where $\alpha=-2.35$).
We performed a Kolomogorov-Smirnov (KS) test,
assuming a single power law distribution. We found values
of $\alpha$ outside the range $-2.7 \ge \alpha \ge -4.4$
inconsistent with the measured distribution
at 95\% confidence. 
Alternatively, we may consider a model distribution that is a
Salpeter IMF ($\alpha=-2.35$) up to some maximum mass, which we may
vary. We found that this model
was inconsistent with the data at 95\% confidence
at assumed maximum masses $> 26$~\msun. 
However, precise determination of this value is difficult in our survey due
to the intrinsic rarity of massive stars and the mass spacing in our
isochrines (see Fig.~\ref{massplot}). Rather, this value represents
the sort of mass range in which one must consider CCSN possible
in order to maintain a Salpeter IMF.

In either scenario, the full distribution of measured
masses suggests that some
fraction of massive stars are not exploding as CCSN.
This result has interesting implications
for CCSN physics. A wide variety of SN channels have been explored both
theoretically and observationally in the literature. Theoretical predictions
have explored the possibility of direct black hole formation
beyond a certain mass threshold
somewhere around $\sim$25 \msun. The manifestation of such events in an overall
mass distribution would be an observed lack of progenitors beyond
the mass threshold, essentially a more bottom-heavy IMF than that of all
massive stars.
While the reality is likely something more complicated
than a well-defined threshold between CCSN and black hole
formation (many different scenarios likely combine
to produce a complicated mass distribution), the qualitative
effect will be that which we observe in our distribution.

\citet{smartt2009} first identified the red supergiant problem,
an observed lack of Type IIP progenitors between 16 and 30 \msun. Many solutions
have been proposed to explain the problem (see \citet{walmswell2012}
and references therein). In addition, the recent SN 2012aw may fall in this mass range
\citep{fraser2012,vandyk2012b}, suggesting the possibility that Type IIP
progenitors do exist in this range and have simply not yet been
observed in sufficient number. We identify six progenitors with median
progenitor masses between 16 and 30~\msun, although
the uncertainties on many of these are large. If our
progenitors were sampled uniformly from a Salpeter IMF,
we would expect to find $\sim$10 progenitors (20\%) in the mass range
from 16 to 30~\msun. Thus while we don't observe a complete
lack of progenitors in the specified mass range, we do observe fewer than we would
expect given a Salpeter IMF distribution. 

Finally, while we assumed our SNR catalogs constituted a complete, unbiased
sample, the possibility exists that selection effects in the catalogs
lead to the lower end of the progenitor mass distribution being
sampled more heavily. In particular, extremely massive progenitors
are likely associated with strong H II regions, where identification
of SNRs is a more difficult observational task. It is possible
SNRs of this type are systematically undersampled in the survey.

We attempted to quantify both the IMF slope and minimum mass
using Markov Chain Monte Carlo methods, but found that the data could
not produce meaningful constraints beyond those determined by our
more simplistic techniques.
We believe the chief reason for this
is the size of errors due to differential reddening of the fields.
Part of the problem is treatment of differential reddening
as a top-hat distribution. In addition, the inclination of M31 contributes
to these high differential extinction values. The application of this
technique to a less inclined galaxy would be of benefit in this
respect. Finally, the simple addition of more progenitor
mass estimates would allow us to better constrain our analysis.
We are currently performing identical analysis on an additional
$\sim$65 SNR in M33 in pursuit of these final two points.

\subsection{Type Ia Contamination}
Type Ia SNRs coincident with star-forming regions could
in principle affect our mass distribution.
To test the possibility of additional Type Ia contamination
beyond the 11\% observed, we examined Galex FUV fluxes
at the sites of all SNR in our sample. We assumed that sites
with the lowest FUV flux corresponded to possible older Type Ia sites
which happened to be coincident with a small amount
of recent SF. The eight lowest flux SNR (not including the 6 with
no recent SF) were BW-18, BW-69, 1-006, 1-010, 2-024, 2-050,
K891, and K956a. From Table~\ref{tab_results}, these additional
SNRs in general have fewer MS stars and less total SF then most
SNRs in the sample. We found that after removing these SNR, our
observed distribution now ranged from $-2.6 \ge \alpha \ge -4.3$,
which is consistent with our earlier measurement. This suggests
that additional Type Ia contamination has little effect on our
overall result, and still results in an IMF that is steeper than
Salpeter (-2.35).

\subsection{Summary}
Using resolved HST photometry, we have analyzed the stars
surrounding 59 SNR in M31. Using CMD fitting, we
calculate a SFH within a 50 pc radius of each SNR.
We find that 53 of the SNR regions
display significant evidence of recent star formation,
which we use to age-date the progenitor star. The remaining
six regions display no recent star formation, and
we consider them either possible Type Ia candidates or the
result of massive runaway progenitor stars.

We examine the distribution of progenitor masses for
our CCSN candidates and find a lack of massive stars
compared to a standard Salpeter IMF ($dN/dM \propto M^{\alpha}$,
where $\alpha=-2.35$). If a uniform single IMF is assumed,
we find values for $\alpha$ outside the range
$-2.7\ge\alpha\ge-4.4$ inconsistent with the measured distribution
at 95\% confidence. Alternatively, if we consider
a distribution that is a Salpeter IMF up to
some maximum mass, we place an upper limit on the
maximum mass allowed at $M_{Max}\sim26$~\msun.
We also estimate a minimum mass for core collapse of between 7.0 and
7.8~\msun,
which is both greater than the maximum mass for white dwarf collapse
\citep{williams2009} and consistent with direct progenitor
measurements \citep{smartt2009,botticella2012}.

\acknowledgments
Z.G.J. is supported in part by funding from the Mary Gates Endowment.
Z.G.J., B.F.W., and J.J.D. are supported in part by GO-12055.
J.W.M. is supported in part by an NSF Astronomy and Astrophysics Postdoctoral
Fellowship under award AST-0802315. 
This work is based on observations made with the NASA/ESA Hubble Space Telescope,
obtained from the data archive at the Space Telescope Science Institute.
Support for this work was provided by NASA through Hubble Fellowship grant 51273.01 awarded
to K.M.G. by the Space Telescope Science Institute.
STScI is operated by the Association of Universities for Research in
Astronomy, Inc. under NASA contract NAS 5-26555.

\clearpage

\clearpage
\begin{center}
\setlength{\extrarowheight}{3pt}
\begin{longtable}{c c c c c c c c}\\
\caption{List of SNR with Deep 2-Filter HST Data}\tabularnewline
SNR ID & RA (degrees) & DEC (degrees) & HST Field & Project ID & Instrument & Filters w/ 50\% Completeness Limits\\
\endhead

\hline
\multicolumn{7}{c}{\citet{magnier1995} SNR}\\
1-006 &  10.6318 &  41.1005 & POS-33 & 10273 & WFPC-2 & F555W=24.5, F814W=23.4\\
1-008 &  10.7675 &  41.6031 & POS-21 & 10273 & ACS & F555W=26.6, F814W=26.0\\
1-009 &  10.7975 &  41.6256 & POS-23 & 10273 & ACS & F555W=25.6, F814W=25.3\\
1-010 &  10.7979 &  41.4853 & POS-18 & 10273 & WFPC-2 & F555W=24.9, F814W=23.7\\
2-016 &  10.3196 &  40.9554 & G-87 & 6671 & WFPC-2 & F555W=25.2, F814W=23.9\\
2-020 &  10.4508 &  41.1138 & G-104 & 10260 & ACS & F606W=25.6, F814W=23.5\\
2-021 &  10.4773 &  40.7866 & G-119 & 6671 & WFPC-2 & F555W=26.2, F814W=25.0\\
2-024 &  10.5958 &  41.0036 & POS-29 & 10273 & ACS & F555W=25.8, F814W=25.1\\
2-025 &  10.6448 &  40.9688 & POS-29 & 10273 & ACS & F555W=25.9, F814W=25.1\\
2-026 &  10.6698 &  41.0447 & POS-30 & 10273 & ACS & F555W=26.0, F814W=25.2\\
2-028 &  10.7376 &  40.9698 & POS-41 & 10273 & ACS & F555W=25.7, F814W=25.4\\
2-044 &  11.1989 &  41.4654 & B08-F10 & 12075 & ACS & F475W=27.3, F814W=26.0\\
2-046 &  11.2928 &  41.5993 & B12-F17 & 12071 & ACS & F475W=27.5, F814W=26.1\\
2-048 &  11.3094 &  41.6033 & B12-F17 & 12071 & ACS & F475W=27.5, F814W=26.1\\
2-049 &  11.3254 &  41.8683 & B15-F08 & 12056 & ACS & F475W=27.4, F814W=26.0\\
2-050 &  11.3662 &  41.8698 & B15-F07 & 12056 & ACS & F475W=27.4, F814W=26.1\\
\hline
\multicolumn{7}{c}{\citet{williams1995} SNR}\\
BW-11 &  10.2467 &  40.6081 & G-76 & 11081 & WFPC-2 & F606W=26.2, F814W=24.9\\
BW-18 &  10.3988 &  41.1155 & G-104 & 10260 & ACS & F606W=26.1, F814W=24.9\\
BW-19 &  10.5408 &  40.9472 & POS-27 & 10273 & ACS & F555W=26.1, F814W=25.5\\
BW-20 &  10.5433 &  40.8644 & POS-26 & 10273 & ACS & F555W=26.0, F814W=25.0\\
BW-31 &  10.7317 &  40.9956 & POS-41 & 10273 & ACS & F555W=26.4, F814W=25.4\\
BW-32 &  10.7329 &  40.9717 & POS-41 & 10273 & ACS & F555W=25.7, F814W=24.6\\
BW-36 &  10.7725 &  41.3750 & G-205 & 10260 & ACS & F606W=25.5, F814W=24.5\\
BW-39 &  10.7933 &  41.6282 & POS-23 & 10273 & ACS & F555W=25.8, F814W=25.4\\
BW-44 &  10.8779 &  41.6882 & POS-24 & 10273 & ACS & F555W=26.5, F814W=25.0\\
BW-60 &  11.0883 &  41.9018 & B15-F12 & 12056 & ACS & F475W=28.0, F814W=26.8\\
BW-61 &  11.1054 &  41.3501 & B06-F10 & 12105 & ACS & F475W=27.3, F814W=25.9\\
BW-65 &  11.1488 &  41.4227 & B06-F04 & 12105 & ACS & F475W=27.3, F814W=25.9\\
BW-66 & 11.1550 & 41.8666 & B15-F17 & 12056 & ACS & F475W=27.5, F814W=26.2\\
BW-69 &  11.1825 &  41.9645 & B17-F18 & 12059 & ACS & F475W=27.6, F814W=26.5\\
BW-71 &  11.1958 &  41.4886 & B08-F04 & 12075 & ACS & F475W=27.5, F814W=26.0\\
BW-74 &  11.2129 &  41.4847 & B08-F04 & 12075 & ACS & F475W=27.1, F814W=25.8\\
BW-76 &  11.2267 &  41.5121 & B08-F04 & 12075 & ACS & F475W=27.3, F814W=25.9\\
BW-77 &  11.2269 &  41.5306 & B08-F04 & 12075 & ACS & F475W=27.3, F814W=25.9\\
BW-81 &  11.2858 &  41.6101 & B12-F17 & 12071 & ACS & F475W=27.4, F814W=26.1\\
BW-82 &  11.2892 &  41.8523 & B15-F08 & 12056 & ACS & F475W=27.6, F814W=26.3\\
BW-84 &  11.3162 &  41.6561 & B12-F11 & 12071 & ACS & F475W=27.4, F814W=26.0\\
BW-86 &  11.3387 &  41.6668 & B12-F11 & 12071 & ACS & F475W=27.3, F814W=26.0\\
BW-89 &  11.3650 &  41.9036 & B15-F01 & 12056 & ACS & F475W=27.5, F814W=26.3\\
BW-102 &  11.4675 &  42.1618 & B21-F11 & 12055 & ACS & F475W=27.6, F814W=26.8\\
BW-105 &  11.6296 &  41.9886 & B18-F03 & 12108 & ACS & F475W=27.6, F814W=26.6\\
BW-106 & 11.6417 & 42.1804 & B21-F08 & 12055 & ACS & F475W=28.0, F814W=27.0\\
BW-110 &  11.6896 &  42.2183 & B21-F01 & 12055 & ACS & F475W=27.7, F814W=27.0\\
\hline
\multicolumn{7}{c}{\citet{braun1993} SNR}\\
K180 &  10.9186 &  41.1814 & B02-F11 & 12073 & ACS & F475W=27.2, F814W=25.9\\
K376 &  11.0850 &  41.5804 & B09-F14 & 12057 & ACS & F475W=27.1, F814W=25.8\\
K446 &  11.1298 &  41.3572 & B06-F04 & 12105 & ACS & F475W=27.3, F814W=26.0\\
K497 &  11.1562 &  41.4133 & B06-F10 & 12105 & ACS & F475W=27.3, F814W=26.0\\
K516/BW-67 &  11.1699 &  41.4145 & B06-F04 & 12105 & ACS & F475W=27.4, F814W=26.0\\
K525A &  11.1820 &  41.4372 & B08-F10 & 12075 & ACS & F475W=27.3, F814W=26.0\\
K526A &  11.1715 &  41.4653 & B08-F10 & 12075 & ACS & F475W=27.3, F814W=26.0\\
K527A &  11.1834 &  41.4465 & B08-F10 & 12075 & ACS & F475W=27.3, F814W=26.0\\
K574 &  11.2101 &  41.4649 & B08-F10 & 12075 & ACS & F475W=27.4, F814W=26.1\\
K594 &  11.2201 &  41.9161 & B15-F10 & 12056 & ACS & F475W=27.2, F814W=26.0\\
K856A &  11.4313 &  41.9313 & B16-F05 & 12106 & ACS & F475W=27.6, F814W=26.4\\
K891 &  11.5405 &  42.2198 & B21-F04 & 12055 & ACS & F475W=27.6, F814W=26.6\\
K908 &  11.6231 &  41.9685 & B18-F03 & 12108 & ACS & F475W=28.0, F814W=27.0\\
K934/BW-107 &  11.6467 &  42.2266 & B21-F01 & 12055 & ACS & F475W=27.6, F814W=26.9\\
K947/2-047 & 11.6689 & 42.1911 & B21-F07 & 12055 & ACS & F475W=27.6, F814W=26.8\\
K956A &  11.6792 &  42.2171 & B21-F01 & 12055 & ACS & F475W=27.7, F814W=27.0\\ 
\label{tab_snr}
\end{longtable}
\end{center}

\newpage

\setlength{\extrarowheight}{4pt}
\begin{center}
\begin{longtable}{c c c c c c c}
\caption{Progenitor Age and Mass Results}\tabularnewline
\hline
SNR ID & \mzams (\msun)& Age (Myr) & \# MS Stars & Total Stars & Total Mass Formed($10^{2}$\msun) & Additional $dA_{v}$ Applied\\ 
\endhead

BW-74 & 17$^{+25}_{-2}$ & 11$^{+2}_{-7}$ &      825 & 4572 & 169 & 0.3 \\
BW-44 & $\ge10$ & $\le26$ &      644 & 2085 & 3 & 1.5 \\
BW-86 & 9.6$^{+2.9}_{-0.6}$ & 28$^{+4}_{-11}$ &      575 & 4344 & 48 & 0.3 \\
BW-84 & 8.7$^{+2.2}_{-0.6}$ & 34$^{+6}_{-12}$ &      572 & 5158 & 49 & 0.3 \\
K527A & 17$^{+25}_{-8}$ & 11$^{+20}_{-7}$ &      531 & 4925 & 62 & 0.1 \\
2-049 & 7.6$^{+34}_{-0.3}$ & 47$^{+3}_{-42}$ &      517 & 4364 & 60 & 0.6 \\
BW-81 & 12$^{+1}_{-2}$ & 19$^{+4}_{-2}$ &      502 & 5212 & 16 & 0.4 \\
K525A & 20$^{+22}_{-12}$ & 9.4$^{+35}_{-4.4}$ &      501 & 4677 & 121 & 0.2 \\
BW-65 & 8.1$^{+7.6}_{-0.40}$ & 40$^{+4.4}_{-28}$ &      500 & 5263 & 57 & 0.5 \\
BW-77 & 7.6$^{+4.1}_{-0.2}$ & 47$^{+3}_{-27}$ &      473 & 4968 & 31 & 0.2 \\
K934 & $\ge$13 & $\ge$16 &      455 & 1934 & 75 & 0.7 \\
K908 & 8.6$^{+4.9}_{-0.4}$ & 36$^{+5}_{-20}$ &      418 & 3360 & 47 & 0.3 \\
BW-66 & 36$^{+6}_{-27}$ & 5.5$^{+26}_{-0.5}$ &      411 & 3310 & 63 & 1.6 \\
BW-71 & 10$^{+32}_{-3}$ & 25$^{+19}_{-20}$ &      401 & 5281 & 37 & 0.6 \\
BW-31 & 9.9$^{+0.3}_{-0.3}$ & 27$^{+2}_{-2}$ &      397 & 2288 & 23 & 0.3 \\
K446 & 9.0$^{+26}_{-0.4}$ & 32$^{+4}_{-26}$ &      389 & 4921 & 26 & 0.6 \\
K594 & 8.4$^{+0.2}_{-0.2}$ & 38$^{+2}_{-2}$ &      379 & 2537 & 26 & 0.5 \\
BW-106 & 19$^{+2}_{-11}$ & 10$^{+30}_{-1}$ &      372 & 2642 & 64 & 1.3 \\
K497 & 10$^{+7}_{-1}$ & 26$^{+2}_{-15}$ &      360 & 5106 & 26 & 0.2 \\
K856A & 27$^{+3}_{-19}$ & 7$^{+43}_{-1}$ &      356 & 2125 & 54 & 0.3 \\
K376 & 16$^{+1}_{-1}$ & 12$^{+1}_{-1}$ &      350 & 6874 & 55 & 0.7 \\
BW-76 & 8.5$^{+2.5}_{-0.3}$ & 37$^{+3}_{-14}$ &      345 & 5013 & 35 & 0.3 \\
1-008 & 7.6$^{+2}_{-0.3}$ & 46$^{+4}_{-18}$ &      336 & 1895 & 61 & 1.7 \\
BW-61 & 15$^{+20}_{-6}$ & 13$^{+18}_{-8}$ &      303 & 4835 & 106 & 0.8 \\
2-050 & 7.6$^{+4}_{-0.3}$ & 47$^{+3}_{-24}$ &      289 & 4343 & 11 & 0.5 \\
K526A & 8.4$^{+0.2}_{-0.2}$ & 38$^{+2}_{-2}$ &      288 & 4842 & 3 & 0.3 \\
BW-82 & 8.1$^{+8}_{-0.4}$ & 40$^{+5}_{-27}$ &      281 & 4017 & 8 & 0.3 \\
BW-89 & 8.1$^{+7}_{-0.4}$ & 40$^{+5}_{-26}$ &      265 & 4296 & 43 & 0.9 \\
K947 & 13$^{+39}_{-1}$ & 16$^{+2}_{-12}$ &      245 & 1874 & 37 & 0.9 \\
BW-60 & 14$^{+1}_{-2}$ & 15$^{+3}_{-1}$ &      243 & 3277 & 39 & 1.5 \\
BW-105 & $\ge$8 & $\le$44 &      239 & 1945 & 13 & 0.3 \\
BW-11 & 10$^{+10}_{-2}$ & 26$^{+19}_{-17}$ &      235 & 1199 & 105 & 0.9 \\
1-009 & 8.9$^{+0.8}_{-0.3}$ & 33$^{+3}_{-5}$ &      219 & 739 & 36 & 1.2 \\
BW-20 & 7.9$^{+0.3}_{-0.3}$ & 42$^{+3}_{-3}$ &      192 & 1339 & 22 & 1.8 \\
2-044 & 8.4$^{+1.8}_{-1.1}$ & 37$^{+13}_{-12}$ &      186 & 4368 & 13 & 0.2 \\
2-046 & 11$^{+1}_{-1}$ & 21$^{+2}_{-2}$ &      185 & 4445 & 16 & 0.4 \\
2-024 & 16$^{+1}_{-1}$ & 12$^{+1}_{-1}$ &      164 & 1934 & 5 & 0.5 \\
BW-39 & $\ge$11 & $\le$20 &      163 & 844 & 104 & 2.1 \\
K574 & 8.1$^{+6}_{-0.4}$ & 40$^{+5}_{-24}$ &      162 & 4207 & 18 & 0.5 \\
K180 & 8.8$^{+0.2}_{-0.2}$ & 33$^{+2}_{-2}$ &      162 & 3950 & 23 & 0.6 \\
2-048 & 7.7$^{+3.3}_{-0.4}$ & 45$^{+5}_{-23}$ &      160 & 3581 & 33 & 0.7 \\
K891 & $\ge$9 & $\le$36 &      149 & 1441 & 23 & 1.1 \\
2-025 & 8.2$^{+0.4}_{-0.9}$ & 39$^{+11}_{-4}$ &      140 & 1287 & 13 & 0.3 \\
BW-18 & 9.9$^{+0.5}_{-0.5}$ & 27$^{+2}_{-2}$ &      117 & 2165 & 25 & 2.0 \\
K516 & 8.5$^{+3}_{-0.3}$ & 36$^{+4}_{-14}$ &      109 & 2210 & 10 & 0.4 \\
K956A & 8.1$^{+18}_{-0.4}$ & 40$^{+5}_{-33}$ &      103 & 1457 & 22 & 1.1 \\
BW-69 & 24$^{+2}_{-9}$ & 7.5$^{+5}_{-1}$ &      100 & 1467 & 8 & 1.5 \\
BW-32 & 9.3$^{+0.3}_{-0.3}$ & 30$^{+2}_{-2}$ &       91 & 562 & 6 & 0.5 \\
BW-110 & 12$^{+1}_{-2}$ & 20$^{+8}_{-2}$ &       86 & 1391 & 12 & 0.8 \\
BW-102 & 8.5$^{+1.5}_{-0.4}$ & 36$^{+4}_{-11}$ &       73 & 1568 & 31 & 1.5 \\
2-020 & 18$^{+1}_{-2}$ & 11$^{+2}_{-1}$ &       38 & 1137 & 20 & 1.4 \\
1-006 & 11$^{+1}_{-1}$ & 24$^{+5}_{-2}$ &       10 & 601 & 6 & 0.0 \\
1-010 & 12$^{+1}_{-1}$ & 19$^{+2}_{-1}$ &       10 & 335 & 9 & 0.2 \\

\multicolumn{7}{c}{Fields with no recent SF (Probable Type 1a or Runaway Progenitors), not included in distribution}\\
\hline
BW-19 & - & - & 291 & 2520 & 0 & 0.5 \\
BW-36 & - & - & 10 & 4162 & 0 & 1.3 \\
2-026 & - & - & 191 & 2421 & 0 & 0.1 \\
2-021 & - & - & 122 & 1268 & 0 & 0.0 \\
2-028 & - & - & 53 & 374 & 0 & 0.5 \\
2-016 & - & - & 5 & 466 & 0 & 0.6 \\
\label{tab_results}
\end{longtable}
\end{center}
\clearpage


\begin{thebibliography}{70}
\expandafter\ifx\csname natexlab\endcsname\relax\def\natexlab#1{#1}\fi

\bibitem[{{Badenes} {et~al.}(2009){Badenes}, {Harris}, {Zaritsky}, \&
  {Prieto}}]{Badenes2009}
{Badenes}, C., {Harris}, J., {Zaritsky}, D., \& {Prieto}, J.~L. 2009, \apj,
  700, 727

\bibitem[{{Barker} {et~al.}(2007){Barker}, {Sarajedini}, {Geisler}, {Harding},
  \& {Schommer}}]{barker2007}
{Barker}, M.~K., {Sarajedini}, A., {Geisler}, D., {Harding}, P., \& {Schommer},
  R. 2007, \aj, 133, 1138

\bibitem[{{Barth} {et~al.}(1996){Barth}, {van Dyk}, {Filippenko}, {Leibundgut},
  \& {Richmond}}]{barth1996}
{Barth}, A.~J., {van Dyk}, S.~D., {Filippenko}, A.~V., {Leibundgut}, B., \&
  {Richmond}, M.~W. 1996, \aj, 111, 2047

\bibitem[{{Bastian} \& {Goodwin}(2006)}]{bastian2006}
{Bastian}, N., \& {Goodwin}, S.~P. 2006, \mnras, 369, L9

\bibitem[{{Blair} {et~al.}(1982){Blair}, {Kirshner}, \&
  {Chevalier}}]{blair1982}
{Blair}, W.~P., {Kirshner}, R.~P., \& {Chevalier}, R.~A. 1982, \apj, 254, 50

\bibitem[{{Botticella} {et~al.}(2012){Botticella}, {Smartt}, {Kennicutt},
  {Cappellaro}, {Sereno}, \& {Lee}}]{botticella2012}
{Botticella}, M.~T., {Smartt}, S.~J., {Kennicutt}, {et al.} 2012, \aap, 537, A132

\bibitem[{{Braun} \& {Walterbos}(1993)}]{braun1993}
{Braun}, R., \& {Walterbos}, R.~A.~M. 1993, \aaps, 98, 327

\bibitem[{{Cappellaro} {et~al.}(1999){Cappellaro}, {Evans}, \&
  {Turatto}}]{cappellaro1999}
{Cappellaro}, E., {Evans}, R., \& {Turatto}, M. 1999, \aap, 351, 459

\bibitem[{{Crockett} {et~al.}(2008){Crockett}, {Eldridge}, {Smartt},
  {Pastorello}, {Gal-Yam}, {Fox}, {Leonard}, {Kasliwal}, {Mattila}, {Maund},
  {Stephens}, \& {Danziger}}]{crockett2008}
{Crockett}, R.~M., {Eldridge}, J.~J., {Smartt}, S.~J., {et al.} 2008, \mnras,
  391, L5

\bibitem[{{Dalcanton} {et~al.}(2009){Dalcanton}, {Williams}, {Seth}, {Dolphin},
  {Holtzman}, {Rosema}, {Skillman}, {Cole}, {Girardi}, {Gogarten},
  {Karachentsev}, {Olsen}, {Weisz}, {Christensen}, {Freeman}, {Gilbert},
  {Gallart}, {Harris}, {Hodge}, {de Jong}, {Karachentseva}, {Mateo}, {Stetson},
  {Tavarez}, {Zaritsky}, {Governato}, \& {Quinn}}]{dalcanton2009}
{Dalcanton}, J.~J., {Williams}, B.~F., {Seth}, A.~C., {et al.} 2009,
  \apjs, 183, 67

\bibitem[{{Dennefeld} \& {Kunth}(1981)}]{dennefeld1981}
{Dennefeld}, M., \& {Kunth}, D. 1981, \aj, 86, 989

\bibitem[{{Dolphin}(2000)}]{dolphin2000}
{Dolphin}, A.~E. 2000, \pasp, 112, 1383

\bibitem[{{Dolphin}(2002)}]{dolphin2002}
---. 2002, \mnras, 332, 91

\bibitem[{{Efremov}(1991)}]{efremov1991}
{Efremov}, Y.~N. 1991, Soviet Astronomy Letters, 17, 173

\bibitem[{{Eldridge} {et~al.}(2011){Eldridge}, {Langer}, \&
  {Tout}}]{eldridge2011}
{Eldridge}, J.~J., {Langer}, N., \& {Tout}, C.~A. 2011, \mnras, 414, 3501

\bibitem[{{Fraser} {et~al.}(2012){Fraser}, {Maund}, {Smartt}, {Botticella},
  {Dall'Ora}, {Inserra}, {Tomasella}, {Benetti}, {Ciroi}, {Eldridge}, {Ergon},
  {Kotak}, {Mattila}, {Ochner}, {Pastorello}, {Reilly}, {Sollerman},
  {Stephens}, {Taddia}, \& {Valenti}}]{fraser2012}
{Fraser}, M., {Maund}, J.~R., {Smartt}, S.~J., {et al.} 2012, arXiv:1204.1523, Submitted to ApJL

\bibitem[{{Gal-Yam} \& {Leonard}(2009)}]{galyam2009}
{Gal-Yam}, A., \& {Leonard}, D.~C. 2009, \nat, 458, 865

\bibitem[{{Gal-Yam} {et~al.}(2007){Gal-Yam}, {Leonard}, {Fox}, {Cenko},
  {Soderberg}, {Moon}, {Sand}, {Caltech Core Collapse Program}, {Li},
  {Filippenko}, {Aldering}, \& {Copin}}]{galyam2007}
{Gal-Yam}, A., {Leonard}, D.~C., {Fox}, D.~B., {et al.} 2007, \apj, 656, 372

\bibitem[{{Gallart} {et~al.}(2005){Gallart}, {Zoccali}, \&
  {Aparicio}}]{gallart2005}
{Gallart}, C., {Zoccali}, M., \& {Aparicio}, A. 2005, \araa, 43, 387

\bibitem[{{Girardi} {et~al.}(2010){Girardi}, {Williams}, {Gilbert},
  {Rosenfield}, {Dalcanton}, {Marigo}, {Boyer}, {Dolphin}, {Weisz},
  {Melbourne}, {Olsen}, {Seth}, \& {Skillman}}]{girardi2010}
{Girardi}, L., {Williams}, B.~F., {Gilbert}, K.~M., {et al.}
  2010, \apj, 724, 1030

\bibitem[{{Gogarten} {et~al.}(2009{\natexlab{a}}){Gogarten}, {Dalcanton},
  {Murphy}, {Williams}, {Gilbert}, \& {Dolphin}}]{gogarten2009}
{Gogarten}, S.~M., {Dalcanton}, J.~J., {Murphy}, J.~W., {et al.}
2009{\natexlab{a}}, \apj, 703, 300

\bibitem[{{Gogarten} {et~al.}(2009{\natexlab{b}}){Gogarten}, {Dalcanton},
  {Williams}, {Seth}, {Dolphin}, {Weisz}, {Skillman}, {Holtzman}, {Cole},
  {Girardi}, {de Jong}, {Karachentsev}, {Olsen}, \& {Rosema}}]{gogarten2009b}
{Gogarten}, S.~M., {Dalcanton}, J.~J., {Williams}, B.~F., {et al.}
2009{\natexlab{b}}, \apj, 691, 115

\bibitem[{{Harris} \& {Zaritsky}(2009)}]{harris2009}
{Harris}, J., \& {Zaritsky}, D. 2009, \aj, 138, 1243

\bibitem[{{Hendry} {et~al.}(2006){Hendry}, {Smartt}, {Crockett}, {Maund},
  {Gal-Yam}, {Moon}, {Cenko}, {Fox}, {Kudritzki}, {Benn}, \&
  {{\O}stensen}}]{hendry2006}
{Hendry}, M.~A., {Smartt}, S.~J., {Crockett}, R.~M., {et al.} 2006, \mnras, 369, 1303

\bibitem[{Horiuchi {et~al.}(2009)Horiuchi, Beacom, \& Dwek}]{horiuchi2009}
Horiuchi, S., Beacom, J.~F., \& Dwek, E. 2009, Phys. Rev. D, 79, 083013

\bibitem[{{Horiuchi} {et~al.}(2011){Horiuchi}, {Beacom}, {Kochanek}, {Prieto},
  {Stanek}, \& {Thompson}}]{horiuchi2011}
{Horiuchi}, S., {Beacom}, J.~F., {Kochanek}, C.~S., {et al.} 2011, \apj, 738, 154

\bibitem[{{Lada} \& {Lada}(2003)}]{lada2003}
{Lada}, C.~J., \& {Lada}, E.~A. 2003, \araa, 41, 57

\bibitem[{{Li} {et~al.}(2011){Li}, {Leaman}, {Chornock}, {Filippenko},
  {Poznanski}, {Ganeshalingam}, {Wang}, {Modjaz}, {Jha}, {Foley}, \&
  {Smith}}]{li2011}
{Li}, W., {Leaman}, J., {Chornock}, R., {et al.} 2011, \mnras, 412, 1441

\bibitem[{{Li} {et~al.}(2005){Li}, {Van Dyk}, {Filippenko}, \&
  {Cuillandre}}]{li2005}
{Li}, W., {Van Dyk}, S.~D., {Filippenko}, A.~V., \& {Cuillandre}, J. 2005,
  \pasp, 117, 121

\bibitem[{{Li} {et~al.}(2006){Li}, {Van Dyk}, {Filippenko}, {Cuillandre},
  {Jha}, {Bloom}, {Riess}, \& {Livio}}]{li2006}
{Li}, W., {Van Dyk}, S.~D., {Filippenko}, A.~V,
  {et al.} 2006, \apj, 641, 1060

\bibitem[{{Li} {et~al.}(2007){Li}, {Wang}, {Van Dyk}, {Cuillandre}, {Foley}, \&
  {Filippenko}}]{li2007}
{Li}, W., {Wang}, X., {Van Dyk}, S.~D., {et al.} 2007, \apj, 661, 1013

\bibitem[{{Magnier} {et~al.}(1995){Magnier}, {Prins}, {van Paradijs}, {Lewin},
  {Supper}, {Hasinger}, {Pietsch}, \& {Truemper}}]{magnier1995}
{Magnier}, E.~A., {Prins}, S., {van Paradijs}, J., {et al.} 1995, \aaps, 114, 215

\bibitem[{{Ma{\'{\i}}z-Apell{\'a}niz}
  {et~al.}(2004){Ma{\'{\i}}z-Apell{\'a}niz}, {Bond}, {Siegel}, {Lipkin},
  {Maoz}, {Ofek}, \& {Poznanski}}]{maiz2004}
{Ma{\'{\i}}z-Apell{\'a}niz}, J., {Bond}, H.~E., {Siegel}, M.~H., {et al.} 2004, \apjl, 615, L113

\bibitem[Maoz 
\& Mannucci(2012)]{maoz2011} Maoz, D., \& Mannucci, F.\ 2012, PASA, 29, 447 

\bibitem[{{Marigo} {et~al.}(2008){Marigo}, {Girardi}, {Bressan}, {Groenewegen},
  {Silva}, \& {Granato}}]{marigo2008}
{Marigo}, P., {Girardi}, L., {Bressan}, A., {et al.} 2008, \aap, 482, 883

\bibitem[{{Maund} {et~al.}(2011){Maund}, {Fraser}, {Ergon}, {Pastorello},
  {Smartt}, {Sollerman}, {Benetti}, {Botticella}, {Bufano}, {Danziger},
  {Kotak}, {Magill}, {Stephens}, \& {Valenti}}]{maund2011}
{Maund}, J.~R., {Fraser}, M., {Ergon}, M., {et al.} 2011, \apjl, 739, L37

\bibitem[{{Maund} \& {Smartt}(2005)}]{maund2005a}
{Maund}, J.~R., \& {Smartt}, S.~J. 2005, \mnras, 360, 288

\bibitem[{{Maund} {et~al.}(2005){Maund}, {Smartt}, \& {Danziger}}]{maund2005}
{Maund}, J.~R., {Smartt}, S.~J., \& {Danziger}, I.~J. 2005, \mnras, 364, L33

\bibitem[{{Murphy} {et~al.}(2011){Murphy}, {Jennings}, {Williams}, {Dalcanton},
  \& {Dolphin}}]{murphy2011}
{Murphy}, J.~W., {Jennings}, Z.~G., {Williams}, B., {Dalcanton}, J.~J., \&
  {Dolphin}, A.~E. 2011, \apjl, 742, L4

\bibitem[{{Panagia} {et~al.}(2000){Panagia}, {Romaniello}, {Scuderi}, \&
  {Kirshner}}]{panagia2000}
{Panagia}, N., {Romaniello}, M., {Scuderi}, S., \& {Kirshner}, R.~P. 2000,
  \apj, 539, 197

\bibitem[Pietrinferni et al.(2004)]{pietrinferni2004} Pietrinferni, A., 
Cassisi, S., Salaris, M., \& Castelli, F.\ 2004, \apj, 612, 168 

\bibitem[{{Salpeter}(1955)}]{salpeter1955}
{Salpeter}, E.~E. 1955, \apj, 121, 161

\bibitem[{{Sasaki} {et~al.}(2012){Sasaki}, {Pietsch}, {Haberl},
  {Hatzidimitriou}, {Stiele}, {Williams}, {Kong}, \& {Kolb}}]{sasaki2012}
{Sasaki}, M., {Pietsch}, W., {Haberl}, F., {et al.} 2012, \aap, 544, 144

\bibitem[{{Schlegel} {et~al.}(1998){Schlegel}, {Finkbeiner}, \&
  {Davis}}]{schlegel98}
{Schlegel}, D.~J., {Finkbeiner}, D.~P., \& {Davis}, M. 1998, \apj, 500, 525

\bibitem[{{Smartt}(2009)}]{smartt2009b}
{Smartt}, S.~J. 2009, \araa, 47, 63

\bibitem[{{Smartt} {et~al.}(2009){Smartt}, {Eldridge}, {Crockett}, \&
  {Maund}}]{smartt2009}
{Smartt}, S.~J., {Eldridge}, J.~J., {Crockett}, R.~M., \& {Maund}, J.~R. 2009,
  \mnras, 395, 1409

\bibitem[{{Smartt} {et~al.}(2003){Smartt}, {Maund}, {Gilmore}, {Tout},
  {Kilkenny}, \& {Benetti}}]{smartt2003}
{Smartt}, S.~J., {Maund}, J.~R., {Gilmore}, G.~F., {et al.} 2003, \mnras, 343, 735

\bibitem[{{Smartt} {et~al.}(2004){Smartt}, {Maund}, {Hendry}, {Tout},
  {Gilmore}, {Mattila}, \& {Benn}}]{smartt2004}
{Smartt}, S.~J., {Maund}, J.~R., {Hendry}, M.~A., {et al.} 2004, Science, 303, 499

\bibitem[{{Smartt} {et~al.}(2002){Smartt}, {Vreeswijk}, {Ramirez-Ruiz},
  {Gilmore}, {Meikle}, {Ferguson}, \& {Knapen}}]{smartt2002}
{Smartt}, S.~J., {Vreeswijk}, P.~M., {Ramirez-Ruiz}, E., {et al.} 2002, \apjl,
  572, L147

\bibitem[{{Smith} {et~al.}(2011{\natexlab{a}}){Smith}, {Li}, {Filippenko}, \&
  {Chornock}}]{smith2011}
{Smith}, N., {Li}, W., {Filippenko}, A.~V., \& {Chornock}, R.
  2011{\natexlab{a}}, \mnras, 412, 1522

\bibitem[{{Smith} {et~al.}(2011{\natexlab{b}}){Smith}, {Li}, {Miller},
  {Silverman}, {Filippenko}, {Cuillandre}, {Cooper}, {Matheson}, \& {Van
  Dyk}}]{smith2011b}
{Smith}, N., {Li}, W., {Miller}, A.~A., {et al.} 2011{\natexlab{b}}, \apj, 732, 63

\bibitem[{{Stanek} \& {Garnavich}(1998)}]{stanek1998}
{Stanek}, K.~Z., \& {Garnavich}, P.~M. 1998, \apjl, 503, L131+

\bibitem[{{Thompson} {et~al.}(2009){Thompson}, {Prieto}, {Stanek}, {Kistler},
  {Beacom}, \& {Kochanek}}]{thompson2009}
{Thompson}, T.~A., {Prieto}, J.~L., {Stanek}, K.~Z., {et al.} 2009, \apj, 705, 1364

\bibitem[{{Tolstoy} {et~al.}(2009){Tolstoy}, {Hill}, \& {Tosi}}]{tolstoy2009}
{Tolstoy}, E., {Hill}, V., \& {Tosi}, M. 2009, \araa, 47, 371

\bibitem[{{Van Dyk} {et~al.}(2012{\natexlab{a}}){Van Dyk}, {Cenko},
  {Poznanski}, {Arcavi}, {Gal-Yam}, {Filippenko}, {Silverio}, {Stockton},
  {Cuillandre}, {Marcy}, {Howard}, \& {Isaacson}}]{vandyk2012b}
{Van Dyk}, S.~D., {Cenko}, S.~B., {Poznanski}, D., {et al.} 2012{\natexlab{a}}, \apj, 756, 131

\bibitem[{{Van Dyk} {et~al.}(2012{\natexlab{b}}){Van Dyk}, {Davidge},
  {Elias-Rosa}, {Taubenberger}, {Li}, {Levesque}, {Howerton}, {Pignata},
  {Morrell}, {Hamuy}, \& {Filippenko}}]{vandyk2012a}
{Van Dyk}, S.~D., {Davidge}, T.~J., {Elias-Rosa}, N., {et al.} 2012{\natexlab{b}}, \aj, 143, 19

\bibitem[{{Van Dyk} {et~al.}(2011){Van Dyk}, {Li}, {Cenko}, {Kasliwal},
  {Horesh}, {Ofek}, {Kraus}, {Silverman}, {Arcavi}, {Filippenko}, {Gal-Yam},
  {Quimby}, {Kulkarni}, {Yaron}, \& {Polishook}}]{vandyk2011}
{Van Dyk}, S.~D., {Li}, W., {Cenko}, S.~B., {et al.} 2011, \apjl, 741, L28

\bibitem[{{Van Dyk} {et~al.}(2003{\natexlab{a}}){Van Dyk}, {Li}, \&
  {Filippenko}}]{vandyk2003a}
{Van Dyk}, S.~D., {Li}, W., \& {Filippenko}, A.~V. 2003{\natexlab{a}}, \pasp,
  115, 448

\bibitem[{{Van Dyk} {et~al.}(2003{\natexlab{b}}){Van Dyk}, {Li}, \&
  {Filippenko}}]{vandyk2003b}
---. 2003{\natexlab{b}}, \pasp, 115, 1289

\bibitem[{{Van Dyk} {et~al.}(2003{\natexlab{c}}){Van Dyk}, {Li}, \&
  {Filippenko}}]{vandyk2003}
---. 2003{\natexlab{c}}, \pasp, 115, 1289

\bibitem[{{Van Dyk} {et~al.}(1999){Van Dyk}, {Peng}, {Barth}, \&
  {Filippenko}}]{vandyk1999}
{Van Dyk}, S.~D., {Peng}, C.~Y., {Barth}, A.~J., \& {Filippenko}, A.~V. 1999,
  \aj, 118, 2331

\bibitem[{{Vink{\'o}} {et~al.}(2009){Vink{\'o}}, {S{\'a}rneczky}, {Balog},
  {Immler}, {Sugerman}, {Brown}, {Misselt}, {Szab{\'o}}, {Csizmadia}, {Kun},
  {Klagyivik}, {Foley}, {Filippenko}, {Cs{\'a}k}, \& {Kiss}}]{vinko2009}
{Vink{\'o}}, J., {S{\'a}rneczky}, K., {Balog}, Z., {et al.} 2009, \apj, 695, 619

\bibitem[{{Walborn} {et~al.}(1993){Walborn}, {Phillips}, {Walker}, \&
  {Elias}}]{walborn1993}
{Walborn}, N.~R., {Phillips}, M.~M., {Walker}, A.~R., \& {Elias}, J.~H. 1993,
  \pasp, 105, 1240

\bibitem[{{Walmswell} \& {Eldridge}(2012)}]{walmswell2012}
{Walmswell}, J.~J., \& {Eldridge}, J.~J. 2012, \mnras, 419, 2054

\bibitem[{{Wang} {et~al.}(2005){Wang}, {Yang}, {Zhang}, {Ma}, {Zhou}, {Li},
  {Lou}, \& {Li}}]{wang2005}
{Wang}, X., {Yang}, Y., {Zhang}, T., {et al.} 2005, \apjl, 626, L89

\bibitem[{{Weisz} {et~al.}(2011){Weisz}, {Dalcanton}, {Williams}, {Gilbert},
  {Skillman}, {Seth}, {Dolphin}, {McQuinn}, {Gogarten}, {Holtzman}, {Rosema},
  {Cole}, {Karachentsev}, \& {Zaritsky}}]{weisz2011}
{Weisz}, D.~R., {Dalcanton}, J.~J., {Williams}, B.~F., {et al.} 2011, \apj, 739, 5

\bibitem[{{Williams} {et~al.}(2009{\natexlab{a}}){Williams}, {Dalcanton},
  {Dolphin}, {Holtzman}, \& {Sarajedini}}]{williams2009c}
{Williams}, B.~F., {Dalcanton}, J.~J., {Dolphin}, A.~E., {Holtzman}, J., \&
  {Sarajedini}, A. 2009{\natexlab{a}}, \apjl, 695, L15

\bibitem[{{Williams} {et~al.}(2009{\natexlab{b}}){Williams}, {Dalcanton},
  {Seth}, {Weisz}, {Dolphin}, {Skillman}, {Harris}, {Holtzman}, {Girardi}, {de
  Jong}, {Olsen}, {Cole}, {Gallart}, {Gogarten}, {Hidalgo}, {Mateo}, {Rosema},
  {Stetson}, \& {Quinn}}]{williams2009b}
{Williams}, B.~F., {Dalcanton}, J.~J., {Seth}, A.~C., {et al.}
  2009{\natexlab{b}}, \aj, 137, 419

\bibitem[{{Williams} {et~al.}(1995){Williams}, {Schmitt}, \&
  {Winkler}}]{williams1995}
{Williams}, B.~F., {Schmitt}, M.~D., \& {Winkler}, P.~F. 1995, in Bulletin of
  the American Astronomical Society, Vol.~27, American Astronomical Society
  Meeting Abstracts \#186, 883--+

\bibitem[{{Williams} {et~al.}(2009{\natexlab{c}}){Williams}, {Bolte}, \&
  {Koester}}]{williams2009}
{Williams}, K.~A., {Bolte}, M., \& {Koester}, D. 2009{\natexlab{c}}, \apj, 693,
  355

\end{thebibliography}
\end{document}